\documentclass[letterpaper,twocolumn,10pt]{article}
\usepackage{usenix2019_v3}

\usepackage{tikz}
\usepackage{amsmath}
\usepackage{amsfonts}
\usepackage{tabularx, booktabs}
\usepackage{filecontents}
\usepackage{subcaption}
\usepackage[english]{babel}
\usepackage{blindtext}
\usepackage{pifont}
\usepackage{mathtools}
\usepackage{bm}
\usepackage{balance}
\usepackage{enumitem}
\usepackage{algorithm}
\usepackage[noend]{algpseudocode}
\usepackage{pgfplots}
\usepackage{fixltx2e}
\usepackage{xspace}
\usepackage{booktabs}
\usepackage{soul}
\usepackage{pifont}	

\newcommand{\xmark}{\ding{55}}
\usepackage{pdfrender}
\newcommand*{\boldcheckmark}{%
  \textpdfrender{
    TextRenderingMode=FillStroke,
    LineWidth=1pt, % half of the line width is outside the normal glyph
  }{\checkmark}%
}

\usetikzlibrary{chains,shapes.multipart}
\usetikzlibrary{shapes,calc,fit}
\usetikzlibrary{automata,positioning,decorations.pathreplacing}

\newcommand{\sequenceD}{%
	\scalebox{0.75}{%
	\begin{tikzpicture}[baseline=+0.85ex]
		\draw[fill=red!30] ([xshift=60pt]0cm,0cm) rectangle ++(10pt, .5cm)node[pos=.5] {1};		
		\draw[fill=cyan!20] ([xshift=48pt]0cm,0cm) rectangle ++(10pt, .5cm)node[pos=.5] {4};
		\draw[fill=blue!20] ([xshift=36pt]0cm,0cm) rectangle ++(10pt, .5cm)node[pos=.5] {5};
		\draw[fill=orange!30] ([xshift=24pt]0cm,0cm) rectangle ++(10pt, .5cm)node[pos=.5] {2};		
		\draw[fill=red!30] ([xshift=12pt]0cm,0cm) rectangle ++(10pt, .5cm)node[pos=.5] {1};
		\draw[fill=orange!30] (0cm,0cm) rectangle ++(10pt, .5cm)node[pos=.5] {2};
	\end{tikzpicture}%
	}\xspace
}

\newcommand{\pktone}{%
	\scalebox{0.75}{%
	\begin{tikzpicture}[baseline=+0.85ex]
		\draw[fill=red!30] (0cm,0cm) rectangle ++(10pt, .5cm) node[pos=.5] {1};
	\end{tikzpicture}%
	}\xspace
}

\newcommand{\pkttwo}{%
	\scalebox{0.75}{%
	\begin{tikzpicture}[baseline=+0.85ex]
		\draw[fill=orange!30] (0cm,0cm) rectangle ++(10pt, .5cm) node[pos=.5] {2};
	\end{tikzpicture}%
	}\xspace
}

\newcommand{\pktfour}{%
	\scalebox{0.75}{%
	\begin{tikzpicture}[baseline=+0.85ex]
		\draw[fill=cyan!20] (0cm,0cm) rectangle ++(10pt, .5cm) node[pos=.5] {4};
	\end{tikzpicture}%
	}\xspace
}

\newcommand{\pktfive}{%
	\scalebox{0.75}{%
	\begin{tikzpicture}[baseline=+0.85ex]
		\draw[fill=blue!20] (0cm,0cm) rectangle ++(10pt, .5cm) node[pos=.5] {5};
	\end{tikzpicture}%
	}\xspace
}

\newcommand{\onetwofourfive}{%
	\scalebox{0.75}{%
		\begin{tikzpicture}[baseline=+0.85ex]
		\draw[fill=red!30] ([xshift=72pt]0cm,0cm) rectangle ++(10pt, .5cm) node[pos=.5] {1};
		\draw[fill=orange!30] ([xshift=60pt]0cm,0cm) rectangle ++(10pt, .5cm) node[pos=.5] {2};		
		\draw[fill=cyan!20] ([xshift=48pt]0cm,0cm) rectangle ++(10pt, .5cm)node[pos=.5] {4};
		\draw[fill=blue!20] ([xshift=36pt]0cm,0cm) rectangle ++(10pt, .5cm)node[pos=.5] {5};
		\end{tikzpicture}%
	}\xspace
}

\newcommand{\oneonetwotwo}{%
	\scalebox{0.75}{%
		\begin{tikzpicture}[baseline=+0.85ex]
		\draw[fill=red!30] ([xshift=72pt]0cm,0cm) rectangle ++(10pt, .5cm) node[pos=.5] {1};
		\draw[fill=red!30] ([xshift=60pt]0cm,0cm) rectangle ++(10pt, .5cm) node[pos=.5] {1};		
		\draw[fill=orange!30] ([xshift=48pt]0cm,0cm) rectangle ++(10pt, .5cm)node[pos=.5] {2};
		\draw[fill=orange!30] ([xshift=36pt]0cm,0cm) rectangle ++(10pt, .5cm)node[pos=.5] {2};
		\end{tikzpicture}%
	}\xspace
}

\newcommand{\onetwoonetwo}{%
	\scalebox{0.75}{%
		\begin{tikzpicture}[baseline=+0.85ex]
		\draw[fill=red!30] ([xshift=72pt]0cm,0cm) rectangle ++(10pt, .5cm) node[pos=.5] {1};
		\draw[fill=orange!30] ([xshift=60pt]0cm,0cm) rectangle ++(10pt, .5cm) node[pos=.5] {2};		
		\draw[fill=red!30] ([xshift=48pt]0cm,0cm) rectangle ++(10pt, .5cm)node[pos=.5] {1};
		\draw[fill=orange!30] ([xshift=36pt]0cm,0cm) rectangle ++(10pt, .5cm)node[pos=.5] {2};
		\end{tikzpicture}%
	}\xspace
}

\newcommand{\oneonefourfive}{%
	\scalebox{0.75}{%
		\begin{tikzpicture}[baseline=+0.85ex]
		\draw[fill=red!30] ([xshift=72pt]0cm,0cm) rectangle ++(10pt, .5cm) node[pos=.5] {1};
		\draw[fill=red!30] ([xshift=60pt]0cm,0cm) rectangle ++(10pt, .5cm) node[pos=.5] {1};		
		\draw[fill=cyan!20] ([xshift=48pt]0cm,0cm) rectangle ++(10pt, .5cm)node[pos=.5] {4};
		\draw[fill=blue!20] ([xshift=36pt]0cm,0cm) rectangle ++(10pt, .5cm)node[pos=.5] {5};
		\end{tikzpicture}%
	}\xspace
}

\newcommand{\fourfivetwotwo}{%
	\scalebox{0.75}{%
		\begin{tikzpicture}[baseline=+0.85ex]
		\draw[fill=cyan!20] ([xshift=72pt]0cm,0cm) rectangle ++(10pt, .5cm) node[pos=.5] {4};
		\draw[fill=blue!20] ([xshift=60pt]0cm,0cm) rectangle ++(10pt, .5cm) node[pos=.5] {5};		
		\draw[fill=orange!30] ([xshift=48pt]0cm,0cm) rectangle ++(10pt, .5cm)node[pos=.5] {2};
		\draw[fill=orange!30] ([xshift=36pt]0cm,0cm) rectangle ++(10pt, .5cm)node[pos=.5] {2};
		\end{tikzpicture}%
	}\xspace
}
\newcommand{\twotwo}{%
	\scalebox{0.75}{%
		\begin{tikzpicture}[baseline=+0.85ex]
		\draw[fill=orange!30] ([xshift=48pt]0cm,0cm) rectangle ++(10pt, .5cm)node[pos=.5] {2};
		\draw[fill=orange!30] ([xshift=36pt]0cm,0cm) rectangle ++(10pt, .5cm)node[pos=.5] {2};
		\end{tikzpicture}%
	}\xspace
}
\newcommand{\oneone}{%
	\scalebox{0.75}{%
		\begin{tikzpicture}[baseline=+0.85ex]
		\draw[fill=red!30] ([xshift=48pt]0cm,0cm) rectangle ++(10pt, .5cm)node[pos=.5] {1};
		\draw[fill=red!30] ([xshift=36pt]0cm,0cm) rectangle ++(10pt, .5cm)node[pos=.5] {1};
		\end{tikzpicture}%
	}\xspace
}

\newcommand{\mlvec}[1]{\bm{#1}}
\DeclareMathOperator*{\argmin}{arg\,min}

\newcommand{\pfour}{P4\textsubscript{16}\xspace}

\newcommand{\myitem}[1]{\vspace*{0.07in}\noindent\textbf{#1}}

\newcommand{\remove}[1]{}
\newcommand{\name}{\textsf{PACKS}\xspace}

\begin{document}

\newtheorem{definition}{Definition}
\newtheorem{theorem}{Theorem}
\newtheorem{corollary}{Corollary}
\newtheorem{lemma}{Lemma}
\newtheorem{problem}{Problem}

%don't want date printed
\date{}

\title{\textit{Everything Matters} in Programmable Packet Scheduling}

\author{
	{\rm Albert Gran Alcoz}\\
	ETH Z\"urich
	\and
	{\rm Balázs Vass}\\
	BME VIK TMIT  \and
	{\rm Gábor Rétvári}\\
	BME VIK TMIT	\and
	{\rm Laurent Vanbever}\\
	ETH Z\"urich
} 

\maketitle
\begin{abstract}

Programmable packet scheduling allows the deployment of scheduling algorithms into existing switches without need for hardware redesign. Scheduling algorithms are programmed by tagging packets with \emph{ranks}, indicating their desired priority. Programmable schedulers then execute these algorithms by serving packets in the order described in their ranks. 

The ideal programmable scheduler is a Push-In First-Out (PIFO) queue, which achieves perfect packet sorting by pushing packets into arbitrary positions in the queue, while only draining packets from the head. Unfortunately, implementing PIFO queues in hardware is challenging due to the need to arbitrarily sort packets at line rate based on their ranks. 

In the last years, various techniques have been proposed, approximating PIFO behaviors using the resources of existing data planes. While promising, approaches to date \emph{only} approximate \emph{one} of the two characteristic behaviors of a PIFO queue: either its \emph{scheduling behavior}, or its \emph{admission control}.

We introduce \name, the \emph{first} programmable scheduler that fully approximates PIFO queues on \emph{all} their behaviors. \name does so by smartly using a set of strict-priority queues. It uses packet-rank information and queue-occupancy levels at enqueue to decide whether to admit packets to the scheduler, and how to map admitted packets to the different queues.  

We fully implement \name in P4 and evaluate it on real workloads. We show that \name better-approximates PIFO than state-of-the-art approaches and scales. We also show that \name runs at line rate on existing hardware (Intel Tofino).

\end{abstract}
\section{Introduction}
\label{sec:introduction}

Packet scheduling is a classical problem in networking that consists in defining the time and the order in which packets at a given buffer should be drained. Many scheduling algorithms have been proposed, trying to achieve different performance objectives, from minimizing tail packet delays, to ensuring fairness between network tenants, or minimizing flow completion times. Traditionally, deploying these algorithms into hardware took years, since new ASICs were required~\cite{programmability_road_ahead}. 

Just recently, programmable scheduling has been proposed, allowing scheduling algorithms to be deployed to existing hardware~\cite{ups,nosilverbullet,pifo0,pifo1}. Operators synthesize scheduling policies by tagging packets with \emph{ranks}, which indicate their scheduling priority. \emph{Programmable schedulers} process these packets, and schedule them following the order of their ranks. 

While ranking algorithms have been already proposed for many scheduling policies~\cite{pifo1, sp-pifo,afq,ups, pieo, acct}, \emph{implementing} the best programmable scheduler is still an open challenge.

\begin{figure}
	\centering
	\includegraphics[width=0.45\textwidth,keepaspectratio]{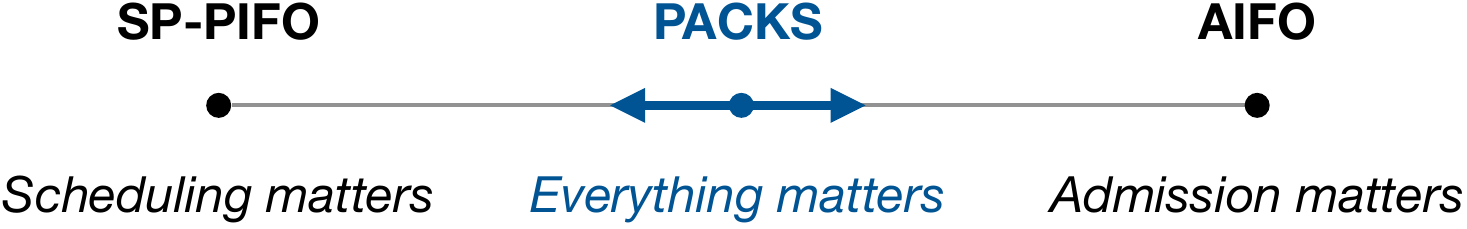}
    \caption{\name navigates the space between SP-PIFO~\cite{sp-pifo} and AIFO~\cite{aifo}, optimizing for both rank ordering and drops.}
	\label{fig:idea}
\end{figure}

Push-In First-Out (PIFO) queues have been proposed as an abstraction for programmable scheduling since they allow to arbitrarily sort packets at line rate based on their ranks~\cite{pifo1}.\footnote{This includes Push-In Extract-Out (PIEO) queues~\cite{pieo}, which we consider an extension of PIFO to support non-work-conserving algorithms (\S\ref{sec:implementation}).} They do so by ``pushing'' packets into arbitrary positions in the queue, while only draining packets from the head. For instance, they can ``push'' incoming packets with low ranks \emph{before} higher-rank packets already in the queue, even dropping the latter if needed to accommodate the new packets. As such, they satisfy the requirements of programmable packet scheduling: (i) they always \emph{admit} the packets with the lowest-ranks, and (ii) they \emph{schedule} packets in perfect order of rank. 

Unfortunately, implementing PIFO queues in hardware is hard, due to the need to sort packets at line rate based on their ranks (even after enqueue), and the need to potentially drop high-rank packets after they have been enqueued (e.g., if a low-rank packet arrives). Recent works have tried to overcome this limitation by \emph{approximating} PIFO behaviors on top of existing programmable data planes~\cite{sp-pifo, afq, pcq, qcluster,aifo,gearbox}. While promising, approaches to date are limited, since they \emph{only} approximate \emph{one} of the two key PIFO characteristics (Fig.~\ref{fig:idea}).

For example, SP-PIFO~\cite{sp-pifo}, QCluster~\cite{qcluster}, AFQ~\cite{afq}, PCQ~\cite{pcq}, and Gearbox~\cite{gearbox}, only focus on approximating PIFO's \emph{scheduling behavior}. They map incoming packets to priority queues to minimize the rank inversions at the output. However, they do so without actively controlling \emph{packet drops}, which they leave as a byproduct effect of the schedulers' design. As such, even though these schedulers can support a broad variety of scheduling algorithms, their behavior can have a negative impact for loss-sensitive applications (cf.~\S\ref{sec:background}). 

On the other hand, AIFO~\cite{aifo} only focuses on approximating PIFO's \emph{admission behavior}. To do so, it executes a rank-aware admission-control policy on top of a single FIFO queue that drops incoming packets similarly to a PIFO queue. Since it runs on a single FIFO queue though, AIFO cannot prioritize traffic according to the packet ranks, which limits the scheduling algorithms that it can approximate accurately.

\begin{table}
	\begin{tabularx}{0.47\textwidth}{p{0.16\textwidth}XXX} \toprule
		& \textsf{SP-PIFO} & \textsf{AIFO} & \textbf{\name} \\
		\midrule
		Scheduling behavior & ~~~~\textcolor{teal}{\boldcheckmark} & ~~~~\textcolor{red}{\xmark} & ~~~~\textcolor{teal}{\boldcheckmark} \\
		Admission behavior & ~~~~\textcolor{red}{\xmark} & ~~~~\textcolor{teal}{\boldcheckmark} & ~~~~\textcolor{teal}{\boldcheckmark} \\
		\bottomrule
	\end{tabularx}
	\caption{PIFO behaviors approximated by existing schedulers.}
	\label{Table: PIFO behaviors}
\end{table}

\myitem{Our work} We propose \name, the \emph{first} programmable packet scheduler that approximates \emph{both}, the admission and scheduling behaviors of a PIFO queue on programmable hardware. \name runs on top of a set of strict-priority queues, and it combines an \emph{admission-control mechanism} and a \emph{queue-mapping technique}. Since \name can \emph{not} drop, nor modify the order of already-enqueued packets (as PIFO queues do), it estimates the expected distribution of incoming packet ranks, it predicts the admission and scheduling behaviors that the PIFO queue would follow, and it executes them \emph{at enqueue}. As such, by approximating all the dimensions of PIFO behaviors, \name satisfies the original vision of programmable-scheduling~\cite{nosilverbullet}: the quest for a \emph{single} scheduling abstraction that can be used to program all types of scheduling algorithms.

\myitem{Evaluation} We fully implement \name in P4 and evaluate it on real workloads. Our results show that \name reduces the scheduling errors by up to 75\% with respect to SP-PIFO, and that it consistently outperforms AIFO in approximating PIFO's admission behavior across various rank distributions. 

\myitem{Contributions} Our main contributions are:
\begin{itemize}
	\itemsep=0pt
	\item \name, a programmable scheduler that emulates PIFO queues on top of a set of strict-priority queues (\S\ref{sec:overview}). 
	\item An admission-control algorithm and a queue-mapping technique, that approximate the PIFO behavior (\S\ref{sec:design1}).
	\item An implementation \footnote{Available at \url{https://github.com/nsg-ethz/packs}} of \name in Java and P4 (\S\ref{sec:implementation}). 
	\item An evaluation showing \name's effectiveness in approximating PIFO using simulations and hardware (\S\ref{sec:evaluation}). 
\end{itemize}
\section{Background}
\label{sec:background}

In this section, we introduce the two programmable packet schedulers that ground the foundations of PIFO's scheduling- and admission-behavior approximation, respectively (Table~\ref{Table: PIFO behaviors}): SP-PIFO~\cite{sp-pifo} (\S\ref{ssec:sp-pifo}) and AIFO~\cite{aifo} (\S\ref{ssec:aifo}). Then, we analyze their limitations and motivate the need for \name (\S\ref{ssec:limitations}). 

\subsection{SP-PIFO}
\label{ssec:sp-pifo}

SP-PIFO~\cite{sp-pifo} approximates PIFO's \emph{scheduling behavior} (i.e., forwarding the earliest-arrived lowest-rank packet first) on a set of strict-priority queues. It does so by dynamically adapting the mapping between packet ranks and priority queues so as to minimize the number of times a higher-rank packet is scheduled before a lower-rank packet in the queue.

\myitem{Mapping} SP-PIFO maps incoming packets to queues based on the queue \emph{bounds}, which define the lowest rank that can be enqueued to each queue. Whenever SP-PIFO receives a packet, it scans the queue bounds bottom-up (i.e., from lowest- to highest-priority queue), and maps the packet to the first queue with a bound lower or equal than the packet rank.

\myitem{Adaptation} SP-PIFO adapts queue bounds dynamically using two mechanisms: a \emph{push-up} stage where future low-rank (i.e., high-priority) packets are pushed to higher-priority queues; and a \emph{push-down} stage where future high-rank (i.e., low-priority) packets are pushed to lower-priority queues. The push-up stage consists in updating each queue bound to the rank of each enqueued packet, every time SP-PIFO enqueues a packet. The push-down stage consists in decreasing the queue bound of \emph{all} queues whenever SP-PIFO  detects a scheduling inversion in the highest-priority queue. With these two stages, SP-PIFO spreads packet ranks across queues, minimizing inversions and approximating PIFO's scheduling behavior.

\subsection{AIFO}
\label{ssec:aifo}

AIFO~\cite{aifo} approximates PIFO's \emph{admission behavior} (i.e., only enqueue the earliest-arrived lowest-rank packets) on a FIFO queue. It does so by maintaining a sliding window of the most-recent packet ranks, and by dropping incoming packets based on their rank and the queue-occupancy level. 

\myitem{Admission} AIFO admits packets based on two dimensions: the distance of the packet rank to the ranks of packets already in the queue, and the time-discrepancy between the arrival rate and the forwarding rate of the FIFO queue. The further the packet rank is from the ranks of recently-enqueued packets, the higher the chances it will be dropped (i.e., AIFO prefers to drop high-rank packets over low-rank packets). Similarly, the lesser space is available in the FIFO queue, the higher the chances that the packet will be dropped. Little space in the queue indicates that the arrival rate is exceeding the departure rate, in which case AIFO drops packets more aggressively.

\subsection{Limitations}
\label{ssec:limitations}

\begin{figure}
	\centering
	\resizebox {0.75\columnwidth} {!} {
		\tikzset{
  queue/.pic={
    \draw[line width=1pt] (0,0) -- ++(1.8cm,0) -- ++(0,-.75cm) -- ++(-1.8cm,0);
  },
  shortqueue/.pic={
    \draw[line width=1pt] (0.8,0) -- ++(1cm,0) -- ++(0,-.75cm) -- ++(-1cm,0);
  }
}

\begin{tikzpicture}[>=latex]
 
  %%%%%%%%%%%%%%%%%%%%%%%%%%%%%%%
  % PIFO
  %%%%%%%%%%%%%%%%%%%%%%%%%%%%%%%
  
  %% Queues
  \path 
  (0.35,1.622cm) pic {queue=1};
  \node at (1.3,0.5cm) {\textsf{PIFO}};

  %% Packets in queue 1
  \draw[fill=orange!30] ([xshift=12pt]0cm,1cm) rectangle ++(10pt, .5cm) node[pos=.5] {2};
  \draw[fill=orange!30] ([xshift=24pt]0cm,1cm) rectangle ++(10pt, .5cm) node[pos=.5] {2};
  \draw[fill=red!30]    ([xshift=36pt]0cm,1cm) rectangle ++(10pt, .5cm) node[pos=.5] {1};
  \draw[fill=red!30]    ([xshift=48pt]0cm,1cm) rectangle ++(10pt, .5cm) node[pos=.5] {1};

  %% Incoming packets
  \draw[fill=orange!30] ([xshift=-60pt]-2cm,1cm) rectangle ++(10pt, .5cm) node[pos=.5] {2};
  \draw[fill=red!30]    ([xshift=-48pt]-2cm,1cm) rectangle ++(10pt, .5cm) node[pos=.5] {1};
  \draw[fill=orange!30] ([xshift=-36pt]-2cm,1cm) rectangle ++(10pt, .5cm) node[pos=.5] {2};
  \draw[fill=blue!20]   ([xshift=-24pt]-2cm,1cm) rectangle ++(10pt, .5cm) node[pos=.5] {5};
  \draw[fill=cyan!20]   ([xshift=-12pt]-2cm,1cm) rectangle ++(10pt, .5cm) node[pos=.5] {4};
  \draw[fill=red!30]    ([xshift=0pt]  -2cm,1cm) rectangle ++(10pt, .5cm) node[pos=.5] (incoming){1};

  %% Outgoing packets
  \draw[fill=red!30]    ([xshift=36pt]3cm,1cm) rectangle ++(10pt, .5cm) node[pos=.5] {1};
  \draw[fill=red!30]    ([xshift=24pt]3cm,1cm) rectangle ++(10pt, .5cm) node[pos=.5] {1};
  \draw[fill=orange!30] ([xshift=12pt]3cm,1cm) rectangle ++(10pt, .5cm) node[pos=.5] {2};
  \draw[fill=orange!30] ([xshift=0pt]3cm,1cm) rectangle ++(10pt, .5cm)  node[pos=.5] (outgoing){2};

  %% Arrows
  \draw[->]  ([xshift=0.25cm] incoming.east) -- ([xshift=2cm] incoming.east);
  \draw[->]  ([xshift=-0.8cm] outgoing.west) -- ([xshift=-0.25cm] outgoing.west);

  %%%%%%%%%%%%%%%%%%%%%%%%%%%%%%%
  % SP-PIFO
  %%%%%%%%%%%%%%%%%%%%%%%%%%%%%%%

    %% Queues
    \path 
    (0.35,-0.378cm) pic {shortqueue=1};
    \path 
    (0.35,-1.3cm) pic {shortqueue=1};
    \node at (1.3,-2.422cm) {\textsf{SP-PIFO}};
  
    %% Packets in queue 1
    \draw[fill=red!30]    ([xshift=36pt]0cm,-1cm) rectangle ++(10pt, .5cm) node[pos=.5] {1};
    \draw[fill=red!30]    ([xshift=48pt]0cm,-1cm) rectangle ++(10pt, .5cm) node[pos=.5] {1};
  
    %% Packets in queue 2
    \draw[fill=blue!20] ([xshift=36pt]0cm,-1.92cm) rectangle ++(10pt, .5cm) node[pos=.5] {5};
    \draw[fill=cyan!20] ([xshift=48pt]0cm,-1.92cm) rectangle ++(10pt, .5cm) node[pos=.5] {4};

    %% Incoming packets
    \draw[fill=orange!30] ([xshift=-60pt]-2cm,-1.5cm) rectangle ++(10pt, .5cm) node[pos=.5] {2};
    \draw[fill=red!30]    ([xshift=-48pt]-2cm,-1.5cm) rectangle ++(10pt, .5cm) node[pos=.5] {1};
    \draw[fill=orange!30] ([xshift=-36pt]-2cm,-1.5cm) rectangle ++(10pt, .5cm) node[pos=.5] {2};
    \draw[fill=blue!20]   ([xshift=-24pt]-2cm,-1.5cm) rectangle ++(10pt, .5cm) node[pos=.5] {5};
    \draw[fill=cyan!20]   ([xshift=-12pt]-2cm,-1.5cm) rectangle ++(10pt, .5cm) node[pos=.5] {4};
    \draw[fill=red!30]    ([xshift=0pt]  -2cm,-1.5cm) rectangle ++(10pt, .5cm) node[pos=.5] (incoming){1};
  
    %% Outgoing packets
    \draw[fill=red!30]    ([xshift=36pt]3cm,-1.5cm) rectangle ++(10pt, .5cm) node[pos=.5] {1};
    \draw[fill=red!30]    ([xshift=24pt]3cm,-1.5cm) rectangle ++(10pt, .5cm) node[pos=.5] {1};
    \draw[fill=cyan!20]   ([xshift=12pt]3cm,-1.5cm) rectangle ++(10pt, .5cm) node[pos=.5] {4};
    \draw[fill=blue!20]   ([xshift=0pt]3cm,-1.5cm) rectangle ++(10pt, .5cm)  node[pos=.5] (outgoing){5};
  
    %% Arrows
    \draw[-]  ([xshift=0.25cm] incoming.east) -- ([xshift=1cm] incoming.east);
    \draw[->]  ([xshift=1cm] incoming.east) |-  ([xshift=2cm,yshift=-0.5cm] incoming.east);
    \draw[->]  ([xshift=1cm] incoming.east) |-  ([xshift=2cm,yshift=0.5cm] incoming.east);

    \draw[->]  ([xshift=-0.8cm, yshift=0.5cm] outgoing.west) -- ([xshift=-0.25cm, yshift=0.5cm] outgoing.west);
    \draw[->]  ([xshift=-0.8cm, yshift=-0.5cm] outgoing.west) -- ([xshift=-0.25cm, yshift=-0.5cm] outgoing.west);

    %% Queue bounds
    \node[above, rectangle, draw] at (0.76cm,-1cm) {$1$};
    \node[above, rectangle, draw] at (0.76cm,-1.92cm) {$2$};

    %%%%%%%%%%%%%%%%%%%%%%%%%%%%%%%
    % AIFO
    %%%%%%%%%%%%%%%%%%%%%%%%%%%%%%%
    
    %% Queues
    \path 
    (0.35,-3.37cm) pic {queue=1};
    \node at (1.3,-4.5cm) {\textsf{AIFO}};

    %% Packets in queue 1
    \draw[fill=orange!30] ([xshift=12pt]0cm,-4cm) rectangle ++(10pt, .5cm) node[pos=.5] {2};
    \draw[fill=red!30]    ([xshift=24pt]0cm,-4cm) rectangle ++(10pt, .5cm) node[pos=.5] {1};
    \draw[fill=orange!30] ([xshift=36pt]0cm,-4cm) rectangle ++(10pt, .5cm) node[pos=.5] {2};
    \draw[fill=red!30]    ([xshift=48pt]0cm,-4cm) rectangle ++(10pt, .5cm) node[pos=.5] {1};

    %% Incoming packets
    \draw[fill=orange!30] ([xshift=-60pt]-2cm,-4cm) rectangle ++(10pt, .5cm) node[pos=.5] {2};
    \draw[fill=red!30]    ([xshift=-48pt]-2cm,-4cm) rectangle ++(10pt, .5cm) node[pos=.5] {1};
    \draw[fill=orange!30] ([xshift=-36pt]-2cm,-4cm) rectangle ++(10pt, .5cm) node[pos=.5] {2};
    \draw[fill=blue!20]   ([xshift=-24pt]-2cm,-4cm) rectangle ++(10pt, .5cm) node[pos=.5] {5};
    \draw[fill=cyan!20]   ([xshift=-12pt]-2cm,-4cm) rectangle ++(10pt, .5cm) node[pos=.5] {4};
    \draw[fill=red!30]    ([xshift=0pt]  -2cm,-4cm) rectangle ++(10pt, .5cm) node[pos=.5] (incoming){1};

    %% Outgoing packets
    \draw[fill=red!30]      ([xshift=36pt]3cm,-4cm) rectangle ++(10pt, .5cm) node[pos=.5] {1};
    \draw[fill=orange!30]   ([xshift=24pt]3cm,-4cm) rectangle ++(10pt, .5cm) node[pos=.5] {2};
    \draw[fill=red!30]      ([xshift=12pt]3cm,-4cm) rectangle ++(10pt, .5cm) node[pos=.5] {1};
    \draw[fill=orange!30]   ([xshift=0pt]3cm,-4cm) rectangle ++(10pt, .5cm)  node[pos=.5] (outgoing){2};

    %% Arrows
    \draw[->]  ([xshift=0.25cm] incoming.east) -- ([xshift=2cm] incoming.east);
    \draw[->]  ([xshift=-0.8cm] outgoing.west) -- ([xshift=-0.25cm] outgoing.west);
    
    %% Admission control
    \node[above, rectangle, draw, fill=white] at (-.6cm,-4cm) {$r<3$};

\end{tikzpicture}
	}
	\caption{SP-PIFO and AIFO can not fully approximate PIFO.}
	\label{fig:example1}
\end{figure}
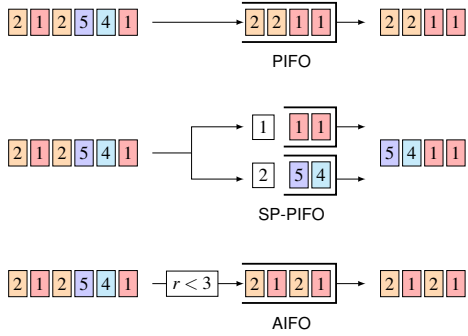

We analyze the limitations of existing schedulers and motivate the need for \name with an example and a simple experiment.

\myitem{Example} Fig.~\ref{fig:example1} shows how PIFO, SP-PIFO and AIFO schedule the packet sequence \sequenceD. We write the first packet received on the far-right (\pktone) and the last, on the far-left (\pkttwo). %Lower ranks indicate higher priority. 
All schedulers have a buffer with capacity for 4 packets, empty at start. SP-PIFO has two priority queues of two packets each. Its queue bounds are fixed, and have values of $1$ and $2$ for the highest- (resp. lowest-) priority queue. AIFO has a fixed admission control that admits packets with rank $r < 3$. 

PIFO ``pushes'' the first four packets into the queue following their rank order: \onetwofourfive. When the fifth packet arrives (\pktone), PIFO ``pushes'' it into the queue between packets with ranks $1$ and $2$, while dropping the highest-rank packet in the queue (\pktfive). When the last packet arrives (\pkttwo), PIFO ``pushes'' it between packets of rank $2$ and $4$, while dropping the packet \pktfour. As a result, the sequence at the output of PIFO is \oneonetwotwo.

SP-PIFO maps packets \oneone to the highest-priority queue, and packets \fourfivetwotwo to the lowest-priority queue (c.f., \S\ref{ssec:sp-pifo}). Since the lowest-priority queue only has room for two packets, the latest-enqueued packets (\twotwo) are dropped. The output sequence is \oneonefourfive, which has sorted ranks (approximating PIFO's scheduling), but does not contain the packets with rank $2$ accepted by PIFO (not approximating PIFO's admission).

AIFO admits the packets with rank $r<3$, same as PIFO. However, since it runs on top of a FIFO queue, it does not prioritize any packet, which results in an output sequence not sorted by rank (i.e., \onetwoonetwo instead of \oneonetwotwo).

\begin{figure}
	\centering
	\begin{subfigure}{0.2356\textwidth}
		\centering
		\includegraphics[width=\textwidth,keepaspectratio]{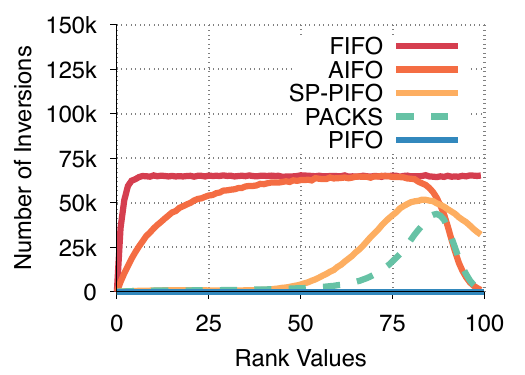}
		\subcaption{Scheduling inversions}
		\label{fig:limitations_uniform_a}
	\end{subfigure}
	\begin{subfigure}{0.2356\textwidth}
		\centering
		\includegraphics[width=\textwidth,keepaspectratio]{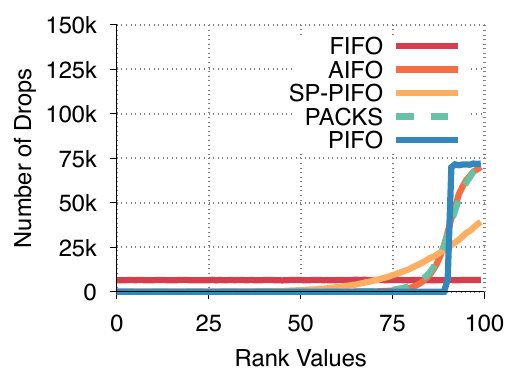}
		\subcaption{Packet drops}
		\label{fig:limitations_uniform_c}
	\end{subfigure}
	\caption{Scheduling performance, uniform rank distribution.}
	\label{fig:limitations_uniform}
\end{figure}

\myitem{Experiment} We now illustrate how the limitations generalize across ranks with an experiment. We implement SP-PIFO, AIFO, \name and FIFO in Netbench~\cite{netbench,Kassing:2017:BFW:3098822.3098836}, a packet-level simulator. We analyze their performance when they schedule a stream of packets with uniformly-distributed ranks across $[0-100]$, over a bottleneck link (details in \S\ref{sec:evaluation}). We measure the scheduling inversions generated by each rank (i.e., the times a scheduled packet has a higher rank than any packet in the queue) and the number of packets dropped of each rank.

Fig.~\ref{fig:limitations_uniform_a} depicts the scheduling inversions across ranks for the different schedulers. PIFO produces no inversions at all, by scheduling packets in perfect order. SP-PIFO approximates this behavior, especially for lower-rank packets, by mapping packets with lower ranks to higher-priority queues. Instead, AIFO and FIFO generate a high number of inversions uniformly across most ranks due to running on a single queue and not being able to prioritize lower-rank packets.

Fig.~\ref{fig:limitations_uniform_c} shows the number of packet drops across ranks for the different schedulers. PIFO performs best, only dropping packets with the highest ranks, since it prioritizes low-rank packets. AIFO closely approximates PIFO's behavior by pro-actively dropping the highest-rank packets. SP-PIFO performs poorly, by just leaving drops as a byproduct effect of its design (i.e., higher-rank packets are dropped more often just because they are mapped to lower-priority queues, which drain less frequently). FIFO performs worst, generating packet drops across all ranks, due to its tail-drop admission fashion.

\name achieves the closest-to-PIFO behavior in both scheduling inversions and packet drops, by combining the best of the two worlds: an admission control like AIFO and a strict-priority queue scheme similar to SP-PIFO.
\section{Overview}
\label{sec:overview}

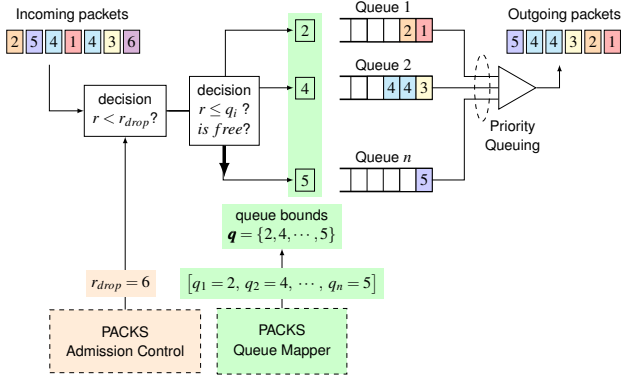
\begin{figure}
	\centering
	\resizebox {1\columnwidth} {!} {
		\tikzset{
  queuei/.pic={
    \draw[line width=1pt]
      (0,0) -- ++(2cm,0) -- ++(0,-.5cm) -- ++(-2cm,0);
    \foreach \Val in {1,...,5}
      \draw ([xshift=-\Val*10pt]2cm,0) -- ++(0,-.5cm);
    \node[above] at (0.95cm,0) {\textsf{Queue} $#1$};
  },
  mytri/.style={
    draw,
    shape=isosceles triangle,
    isosceles triangle apex angle=60,
    inner xsep=0.3cm
  }
}

\begin{tikzpicture}[>=latex]
  
  %% Queues
  \path 
  (0.25,1.5cm) pic {queuei=1}
  (0.25,.25cm) pic {queuei=2}
  (0.25,-1.75cm) pic {queuei=n};

  %% Packets in queue 1
  \draw[fill=white] ([xshift=20pt].5cm,1cm) rectangle ++(10pt, .5cm) node[pos=.5] {};
  \draw[fill=white] ([xshift=10pt].5cm,1cm) rectangle ++(10pt, .5cm) node[pos=.5] {};
  \draw[fill=orange!30] ([xshift=30pt].5cm,1cm) rectangle ++(10pt, .5cm) node[pos=.5] {2};
  \draw[fill=red!30] ([xshift=40pt].5cm,1cm) rectangle ++(10pt, .5cm) node[pos=.5] {1};

  %% Packets in queue 2
  \draw[fill=white] ([xshift=10pt].5cm,-0.25cm) rectangle ++(10pt, .5cm) node[pos=.5] {};
  \draw[fill=cyan!20] ([xshift=20pt].5cm,-0.25cm) rectangle ++(10pt, .5cm) node[pos=.5] {4};
  \draw[fill=cyan!20] ([xshift=30pt].5cm,-0.25cm) rectangle ++(10pt, .5cm)
  node[pos=.5] {4};
  \draw[fill=yellow!20] ([xshift=40pt].5cm,-0.25cm) rectangle ++(10pt, .5cm)
  node[pos=.5] {3};

  %% Packets in queue n
  \draw[fill=white] ([xshift=30pt].5cm,-2.25cm) rectangle ++(10pt, .5cm) node[pos=.5] {};
  \draw[fill=blue!20] ([xshift=40pt].5cm,-2.25cm) rectangle ++(10pt, .5cm)
  node[pos=.5] {5};

  %% Admission
  \node[draw, dashed, align=center,inner sep=10pt, fill=orange!15]
    at (-4.4,-5.5) (ad)
    {\textsf{PACKS} \\ \textsf{Admission Control}};

  \node[draw,align=center,inner sep=5pt,fill=white]
    at (-4.4,-.5cm) (admission)
    {\textsf{decision}\\$r < r_{drop}$\textsf{?}};

  %% Queue Mapper
  \node[draw, dashed, align=center,inner sep=10pt, fill=green!20]
    at (-1,-5.5) (mapper)
    {\textsf{PACKS} \\ \textsf{Queue Mapper}};

  %% Queue bound green box
  \path 
    (-.75,1.5cm) coordinate (aux1)  
    (-.25,-2.25cm) coordinate (aux2)
    (-.75,0cm) coordinate (aux3)
    (-.25,0cm) coordinate (aux4);
  \node[draw,fill=green!20,draw=none,text width=2.5cm,fit={(aux1) (aux2) (aux3) (aux4)}] (dashed) {};

  %% Queue bounds
  \node[above, rectangle, draw] at (-.5cm,1cm) {$2$};
  \node[above, rectangle, draw] at (-.5cm,-.25cm) {$4$};
  \node[above, rectangle, draw] at (-.5cm,-2.25cm) {$5$};

  \node[draw,fill=green!20,align=center,draw=none]
    at (-1,-3) (bound)
    {\textsf{queue bounds}\\$\mlvec{q} = \{2, 4, \cdots, 5\}$};

  %% Incoming packets
  \draw[fill=violet!30] ([xshift=48pt]-6.125cm,0.75cm) rectangle ++(10pt, .5cm) node[pos=.5] {6};
  \draw[fill=yellow!20] ([xshift=36pt]-6.125cm,0.75cm) rectangle ++(10pt, .5cm) node[pos=.5] {3};
  \draw[fill=cyan!20] ([xshift=24pt]-6.125cm,0.75cm) rectangle ++(10pt, .5cm) node[pos=.5] {4};
  \draw[fill=red!30] ([xshift=12pt]-6.125cm,0.75cm) rectangle ++(10pt, .5cm) node[pos=.5] {1};
  \draw[fill=cyan!20] ([xshift=0pt]-6.125cm,0.75cm) rectangle ++(10pt, .5cm)node[pos=.5] {4};
  \draw[fill=blue!20] ([xshift=-12pt]-6.125cm,0.75cm) rectangle ++(10pt, .5cm)node[pos=.5] {5};
  \draw[fill=orange!30] ([xshift=-24pt]-6.125cm,0.75cm) rectangle ++(10pt, .5cm)node[pos=.5] {2};

  %% Outgoing packets
  \draw[fill=red!30] ([xshift=48pt]4.30cm,0.75cm) rectangle ++(10pt, .5cm) node[pos=.5] {1};
  \draw[fill=orange!30] ([xshift=36pt]4.30cm,0.75cm) rectangle ++(10pt, .5cm)node[pos=.5] {2};
  \draw[fill=yellow!20] ([xshift=24pt]4.30cm,0.75cm) rectangle ++(10pt, .5cm)node[pos=.5] {3};
  \draw[fill=cyan!20] ([xshift=12pt]4.30cm,0.75cm) rectangle ++(10pt, .5cm)node[pos=.5] {4};
  \draw[fill=cyan!20] ([xshift=0pt]4.30cm,0.75cm) rectangle ++(10pt, .5cm)node[pos=.5] {4};
  \draw[fill=blue!20] ([xshift=-12pt]4.30cm,0.75cm) rectangle ++(10pt, .5cm)node[pos=.5] {5};

  \node[draw,align=center,mytri]
    at (4,0) (multi)
    {};
  \node[draw,ellipse,dashed,minimum height=1.45cm]
    at ([xshift=-10pt]multi.west) (robin)
    {};
  \node[align=center,anchor=south]
    at ([yshift=-1.25cm]multi.south)
    {\textsf{Priority}\\ \textsf{Queuing}};

  \node[align=left,anchor=south]
    at ([yshift=30pt]-5.55cm,0.25cm)
    {\textsf{Incoming packets}};
  \node[align=left,anchor=south]
    at ([yshift=30pt]5.125cm,0.25cm)
    {\textsf{Outgoing packets}};

  %% Arrows
  \draw[->]
    ([xshift=.0475cm]admission.east) --
    ++(35pt,0pt) |- node[above,pos=0.67] {} coordinate (aux5)
    (-.75,1.25);

  \draw[->]
    (admission.east) --
    ++(35pt,0pt) |- node[above,pos=0.67] {} coordinate (aux6)
    (-.75,0);     

    \draw[->]
    (admission.east) --
    ++(35pt,0pt) |- node[above,pos=0.67] {} coordinate (aux7)
    (-.75,-2);

  % Decision
  \node[draw,,align=center,inner sep=5pt,fill=white]
  at (-2.25cm,-.5cm) (decision)
  {\textsf{decision}\\$r \leq q_i$ \textsf{?} \\ $is~free$\textsf{?}};

  \draw[->]
    (mapper.north) -- (bound.south);

  \draw[->]
    (ad.north) -- (admission.south);

  \draw[->]
    ([xshift=-.75cm, yshift=1cm]admission.west) --
    ([xshift=-.75cm]admission.west) -- (admission.west);
  
  \draw
    (2.25,1.25) --
    ++(20pt,0pt) |- 
    ( $ (multi.west)!0.8!(multi.north west) $);     
  
  \draw
    (2.25,0) -- (multi.west);
    
  \draw
    (2.25,-2) --
    ++(20pt,0pt) |- 
    ( $ (multi.west)!0.8!(multi.south west) $);    

  \draw[->] 
    (multi.east) -- 
    ([xshift=.5cm]multi.east) --
    ([xshift=.5cm,yshift=.5cm]multi.east);

  \draw[->,line width=2pt] 
    (decision.south) -- ([yshift=-19.5pt]decision.south);

  %% Queue mapping config
  \node[draw=none,fill=green!20] at (-1, -4.25) {\textsf{$\big[q_1 = 2, \; q_2 = 4, \; \cdots, \; q_n = 5\big]$}};

  %% Rank-drop config
  \node[draw=none,fill=orange!15] at (-4.5, -4.25) {\textsf{$r_{drop} = 6$}};

\end{tikzpicture}
	}
	\caption{Overview of \name data-plane pipeline.}
	\label{fig:packs}
\end{figure}

We now provide an overview of how \name approximates the behavior of a PIFO queue on existing hardware. \name runs on top of a set of strict-priority queues, and incorporates: (i) an \emph{admission-control} mechanism that decides which packets to admit, and (ii) a \emph{queue mapper} that decides how to map admitted packets to the different priority queues (see Fig.~\ref{fig:packs}). 

With this setup, \name approximates two PIFO behaviors: \emph{admitting} the packets with the lowest-ranks, and \emph{scheduling} packets in perfect order of rank. PIFO does so by being able to drop packets after they have been enqueued, and to sort packets based on their ranks after they have been enqueued, which is not supported (by default) in existing hardware queues.

\myitem{Insight} \name approximates these behaviors by \emph{predicting} the distribution of packets that will arrive during a certain scheduling interval, using this information to \emph{anticipate} the admission and scheduling decisions that the PIFO queue would do, and executing those actions \emph{at enqueue}. Specifically, given a monitored rank distribution, \name first predicts the set of packets with the lowest ranks that are expected to fit into the available buffer space, and proactively drops all the arriving packets that have higher rank. Second, \name estimates the set of admitted packets that should be mapped to each priority queue to optimize the rank order at the output of the scheduler.

\myitem{Rank-distribution estimation} \name uses a \emph{sliding window} to estimate the distribution of packet ranks that are expected to arrive. It assumes that the distribution of latest-enqueued packets gives a good estimate of the one of incoming packets.

\myitem{Admission control} Given the monitored distribution, \name estimates which of the expected packets in the distribution should be admitted to the queues. Intuitively, \name should only admit the packets with the lowest rank that can fit in the available buffer space (since this is what a PIFO queue does). As such, whenever a packet arrives, \name measures the remaining buffer space available, as a percentage of the total buffer space, and computes a rank value, $r_{drop}$, that indicates the first rank for which the quantile of the rank distribution exceeds the percentage of the remaining buffer space. This value $r_{drop}$ is the lowest rank that \name should already drop, to make sure that the admitted packets (i.e., those with rank $r<r_{drop}$) can fit in the available buffer space. Thus, after having computed $r_{drop}$, \name admits the incoming packet \emph{only} if its rank is lower than $r_{drop}$. This results in \name admitting the \emph{lowest-rank} packets that are \emph{expected} to arrive and that fit in the available buffer space, same as PIFO does.

\myitem{Queue mapping} \name then leverages the monitored distribution to find the best mapping of expected packets to priority queues to maximize the rank order at the output of the scheduler. Intuitively, the best mapping is the one that maps packets with lower-ranks to the higher-priority queues, and that minimizes the number of different-rank packets that are mapped to the same queue. Similarly to how $r_{drop}$ drives the admission control, \name defines a set of rank values $\mlvec{q}=(q_1, ..., q_n)$ that drive the mapping of packets to priority queues. For each queue $i$, the queue bound $q_i$ identifies the highest rank that should be admitted to the queue such that the admitted packets (i.e., those with rank $r<q_i$) are the ones with the lowest rank that can fit in the available queue space. With this definition, \name maps low-rank packets to high-priority queues by scanning queue bounds top-down (i.e., from highest- to lowest-priority), and mapping each incoming packet to the first queue for which the packet rank is lower or equal than the queue bound. This results in \name prioritizing \emph{expected} packets of low rank over higher-rank ones, similarly to PIFO.

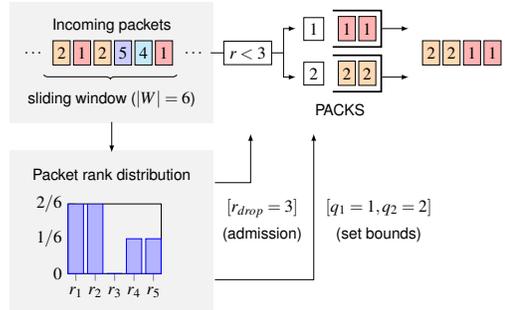
\begin{figure}
	\centering
	\resizebox {0.8\columnwidth} {!} {
		\tikzset{
  queue/.pic={
    \draw[line width=1pt] (0,0) -- ++(1.8cm,0) -- ++(0,-.75cm) -- ++(-1.8cm,0);
  },
  shortqueue/.pic={
    \draw[line width=1pt] (0.8,0) -- ++(1cm,0) -- ++(0,-.75cm) -- ++(-1cm,0);
  }
}

\begin{tikzpicture}[>=latex]

    %%%%%%%%%%%%%%%%%%%%%%%%%%%%%%%
    % PACKS
    %%%%%%%%%%%%%%%%%%%%%%%%%%%%%%%

    %% Queues
    \path 
    (0.35,-0.378cm) pic {shortqueue=1};
    \path 
    (0.35,-1.3cm) pic {shortqueue=1};
    \node at (1.3,-2.422cm) {\textsf{PACKS}};
  
    %% Packets in queue 1
    \draw[fill=red!30]    ([xshift=36pt]0cm,-1cm) rectangle ++(10pt, .5cm) node[pos=.5] {1};
    \draw[fill=red!30]    ([xshift=48pt]0cm,-1cm) rectangle ++(10pt, .5cm) node[pos=.5] {1};
  
    %% Packets in queue 2
    \draw[fill=orange!30] ([xshift=36pt]0cm,-1.92cm) rectangle ++(10pt, .5cm) node[pos=.5] {2};
    \draw[fill=orange!30] ([xshift=48pt]0cm,-1.92cm) rectangle ++(10pt, .5cm) node[pos=.5] {2};

    %% Incoming packets
		\draw[fill=gray!10, draw=none] (-5.5cm,-2.7cm) rectangle ++(4.2cm, 2.5cm) node[pos=.5] (incoming_box){};

		\node[align=left] at (-3.4cm,-0.7cm) {\textsf{Incoming packets}};
		\draw[decorate,decoration={brace,raise=0.1cm}] ([xshift=78pt]-4.8cm,-1.6) -- (-4.8cm,-1.6cm);
		\node[align=left] at (-3.4cm,-2.2cm) {\textsf{sliding window ($|W|=6$)}};

    \draw[fill=orange!30] ([xshift=-60pt]-2.5cm,-1.5cm) rectangle ++(10pt, .5cm) node[pos=.5] {2};
    \draw[fill=red!30]    ([xshift=-48pt]-2.5cm,-1.5cm) rectangle ++(10pt, .5cm) node[pos=.5] {1};
    \draw[fill=orange!30] ([xshift=-36pt]-2.5cm,-1.5cm) rectangle ++(10pt, .5cm) node[pos=.5] {2};
    \draw[fill=blue!20]   ([xshift=-24pt]-2.5cm,-1.5cm) rectangle ++(10pt, .5cm) node[pos=.5] {5};
    \draw[fill=cyan!20]   ([xshift=-12pt]-2.5cm,-1.5cm) rectangle ++(10pt, .5cm) node[pos=.5] {4};
    \draw[fill=red!30]    ([xshift=0pt]  -2.5cm,-1.5cm) rectangle ++(10pt, .5cm) node[pos=.5] (incoming){1};
		
		\node[above, rectangle, draw=none] at (-5cm,-1.4cm) {$\dots$};
		\node[above, rectangle, draw=none] at (-1.7cm,-1.4cm) {$\dots$};
  
    %% Outgoing packets
    \draw[fill=red!30]    ([xshift=36pt]3cm,-1.5cm) rectangle ++(10pt, .5cm) node[pos=.5] {1};
    \draw[fill=red!30]    ([xshift=24pt]3cm,-1.5cm) rectangle ++(10pt, .5cm) node[pos=.5] {1};
    \draw[fill=orange!30] ([xshift=12pt]3cm,-1.5cm) rectangle ++(10pt, .5cm) node[pos=.5] {2};
    \draw[fill=orange!30] ([xshift=0pt]3cm,-1.5cm) rectangle ++(10pt, .5cm)  node[pos=.5] (outgoing){2};
  
    %% Arrows
    \draw[-]   ([xshift=0.75cm] incoming.east) -- ([xshift=2.25cm] incoming.east);
    \draw[->]  ([xshift=2.25cm] incoming.east) |-  ([xshift=2.5cm,yshift=-0.5cm] incoming.east);
    \draw[->]  ([xshift=2.25cm] incoming.east) |-  ([xshift=2.5cm,yshift=0.5cm] incoming.east);

    \draw[->]  ([xshift=-0.8cm, yshift=0.5cm] outgoing.west) -- ([xshift=-0.25cm, yshift=0.5cm] outgoing.west);
    \draw[->]  ([xshift=-0.8cm, yshift=-0.5cm] outgoing.west) -- ([xshift=-0.25cm, yshift=-0.5cm] outgoing.west);

    %% Queue bounds
    \node[above, rectangle, draw] at (0.76cm,-1cm) (bound1) {$1$};
    \node[above, rectangle, draw] at (0.76cm,-1.92cm) (bound2) {$2$};

    %% Admission control
    \node[above, rectangle, draw, fill=white] at (-.6cm,-1.5cm) (admission){$r<3$};
		
		%% Packet-rank distribution
		\draw[fill=gray!10, draw=none] (-5.5cm,-6.6cm) rectangle ++(4.2cm, 3.35cm) node[pos=.5] (distrib_box){};
    \draw[->]  ([xshift=0cm, yshift=-1.1cm] incoming_box.south) -- ([xshift=0cm, yshift=1.5cm] distrib_box.north);
    \draw[->]  ([xshift=2cm,yshift=-1cm] distrib_box.east) -| ([xshift=0cm,yshift=-1cm] bound2.south);
    \draw[->]  ([xshift=2cm,yshift=1cm] distrib_box.east) -| ([xshift=-1.3cm,yshift=-1cm] bound2.south);

		\begin{axis}[
				ybar, 
				ymax=2, 
				width=3.5cm,
				at={(-4.3cm,-5.8cm)},
				bar width=9pt,
				ymin=0,
				title = {},
				ylabel = {},
				xtick=data,
				ytick={0,1,2,3},
				tick pos=left,
				yticklabels={0,$1/6$,$2/6$},
				xticklabels={$r_1$,$r_2$,$r_3$,$r_4$,$r_5$}
				]
			\addplot coordinates {
				 (1, 2) (2, 2) (3, 0) (4, 1) (5, 1) 
			};
		\end{axis}
		\node[draw=none, anchor=south, align=center] at (-3.4cm,-4cm) {\textsf{Packet rank distribution}};
		
    %% Updated bounds
    \node[align=center] at (2.1cm,-4.6cm) {$[q_1=1,q_2=2]$\\};
    \node[align=center] at (2.1cm,-5cm) {\textsf{(set bounds)}};

    \node[align=center] at (-0.3cm,-4.6cm) {$[r_{drop}=3]$\\};
    \node[align=center] at (-0.3cm,-5cm) {\textsf{(admission)}};

\end{tikzpicture}
	}
	\caption{\name closely approximates PIFO's behavior.}
	\label{fig:example1_packs}
\end{figure}

\myitem{Example} Fig.~\ref{fig:example1_packs} shows how \name schedules the sequence \sequenceD. We assume that the sequence is received over and over, and configure \name with two priority queues of two packets and a sliding window of size $|W| = 6$. After receiving the 6-th packet, \name has estimated the rank distribution, where the probability of receiving a packet of ranks 1 to 5 are $p(1)=2/6$, $p(2)=2/6$, $p(3)=0$, $p(4)=1/6$, $p(5)=1/6$. Given the available buffer space (i.e., 4 packets), and based on the monitored rank distribution, \name sets $r_{drop}$ to $3$, since the expected $4$ packets with lowest rank are those with rank $1$ and $2$. Then, \name sets $q_i$ based on the available buffer space at each queue (i.e., $2$ packets each). As such, it sets $q_1 = 1$ to map the two expected packets with lowest rank to the highest-priority queue, and $q_2=2$, to map the two expected admitted packets with highest rank to the lowest-priority queue. As a result, when draining the queues in order of priority, the sequence at the output of \name is \oneonetwotwo, the exact same one as in the PIFO queue (see Fig.~\ref{fig:example1}).
\section{\name design}
\label{sec:design1}

In this section, we describe the theoretical basis supporting the design of \name. First, we frame the problem and introduce the design space (\S\ref{ssec:designspace}). Second, we provide the high-level intuition behind \name's design by studying the case in which it schedules a batch of packets (\S\ref{ssec:intuition}). Third, we generalize the algorithm to the online setup (\S\ref{ssec:onlineadaptation}). Finally, we wrap up the section by summarizing the \name's algorithm (\S\ref{ssec:algorithm}).

\subsection{Design space}
\label{ssec:designspace}

\begin{figure}
	\centering
	\resizebox {1\columnwidth} {!} {
		\tikzset{
  queuei/.pic={
    \draw[line width=1pt]
      (0,0) -- ++(2cm,0) -- ++(0,-.5cm) -- ++(-2cm,0);
    \foreach \Val in {1,...,5}
      \draw ([xshift=-\Val*10pt]2cm,0) -- ++(0,-.5cm);
    \node[above] at (0.95cm,0) {\textsf{Queue} $#1$};
    },
  mytri/.style={
    draw,
    shape=isosceles triangle,
    isosceles triangle apex angle=60,
    inner xsep=0.3cm
    }
}

\begin{tikzpicture}[>=latex]
  
  %% Queues
  \path 
  (0.25,1.5cm) pic {queuei=1}
  (0.25,.25cm) pic {queuei=2}
  (0.25,-1.75cm) pic {queuei=n};

  %% Admission control
  \node[draw,align=center,inner sep=10pt]
    at (-4.4,-.5cm) (admission)
    {Admission \\ Control};

  %% Incoming packets
  \draw[fill=yellow!20] ([xshift=36pt]-6.125cm,0.75cm) rectangle ++(10pt, .5cm) node[pos=.5] {3};
  \draw[fill=cyan!20] ([xshift=24pt]-6.125cm,0.75cm) rectangle ++(10pt, .5cm) node[pos=.5] {4};
  \draw[fill=red!30] ([xshift=12pt]-6.125cm,0.75cm) rectangle ++(10pt, .5cm) node[pos=.5] {1};
  \draw[fill=cyan!20] ([xshift=0pt]-6.125cm,0.75cm) rectangle ++(10pt, .5cm)node[pos=.5] {4};
  \draw[fill=blue!20] ([xshift=-12pt]-6.125cm,0.75cm) rectangle ++(10pt, .5cm)node[pos=.5] {5};
  \draw[fill=orange!30] ([xshift=-24pt]-6.125cm,0.75cm) rectangle ++(10pt, .5cm)node[pos=.5] {2};

  %% Outgoing packets
  \node[draw=none,align=center,text=RoyalBlue] at (5,0.85) {\LARGE ?};

  \node[draw,align=center,mytri] at (4,0) (multi) {};
  \node[draw,ellipse,dashed,minimum height=1.45cm] at ([xshift=-10pt]multi.west) (robin) {};

  \node[align=center,anchor=south] at ([yshift=-1.25cm]multi.south) {\textsf{Priority}\\ \textsf{Queuing}};
  \node[align=left,anchor=south] at ([yshift=30pt]-5.725cm,0.25cm) {\textsf{Incoming packets}};
  \node[align=left,anchor=south] at ([yshift=30pt]5.125cm,0.25cm) {\textsf{Outgoing packets}};

  %% Arrows
  \draw[->]
    (admission.east) --
    ++(30pt,0pt) |- node[above,pos=0.7] {} coordinate (aux5) (0,1.25);

  \draw[->]
    (admission.east) --
    ++(30pt,0pt) |- node[above,pos=0.7] {} coordinate (aux6) (0,0);     

  \draw[->]
    (admission.east) --
    ++(30pt,0pt) |- node[above,pos=0.7] {} coordinate (aux7) (0,-2);

  %% Arrow incoming packets to admission control
  \draw[->]
    ([xshift=-.75cm, yshift=1cm]admission.west) --
    ([xshift=-.75cm]admission.west) -- (admission.west);
  
  %% Arrows after the queues
  \draw
    (2.25,1.25) --
    ++(20pt,0pt) |- 
    ( $ (multi.west)!0.8!(multi.north west) $);     
  
  \draw
    (2.25,0) -- (multi.west);
    
  \draw
    (2.25,-2) --
    ++(20pt,0pt) |- 
    ( $ (multi.west)!0.8!(multi.south west) $);    

  \draw[->] 
    (multi.east) -- 
    ([xshift=.5cm]multi.east) --
    ([xshift=.5cm,yshift=.5cm]multi.east);

  %% Queue mapping
  \node[draw,fill=white,align=center,inner sep=10pt]
  at (-2cm,-.5cm) (mapping)
  {Queue \\ Mapping};

\end{tikzpicture}
	}
	\caption{\name's design space.}
	\label{fig:abstraction}
\end{figure}
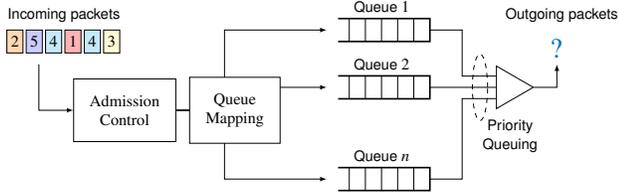

Let us consider the scheduling design space in Fig.~\ref{fig:abstraction}, which represents the available resources in existing data planes. It contains a fixed set of strict-priority queues (of fixed size), and two additional components: an \emph{admission-control} mechanism that decides which packets to admit, and a \emph{queue mapper} that decides how to map admitted packets to the priority queues. \footnote{Some devices allow extra functionalities such as flexible priority-queue configuration, round-robin scheduling, or buffer management. We use Fig.~\ref{fig:abstraction}'s abstraction for generality and to guarantee line-rate processing for \emph{all} packets.} After a packet is mapped to a queue, the packet is enqueued \emph{only} if the queue has enough buffer space to accommodate the new packet; otherwise, the packet is dropped. The scheduler constantly drains queues in decreasing order of priority (i.e., only scheduling packets from low-priority queues when higher-priority queues are empty), and it schedules packets within each priority queue in a first-in first-out fashion.

\myitem{Problem}~~\textit{How can we best approximate the behavior of a PIFO queue on top of the \name abstraction in Fig.~\ref{fig:abstraction}?} 

\noindent The \name abstraction only allows for two design decisions: an admission-control protocol (i.e., which packets should we admit?) and a queue-mapping algorithm (i.e., how should we map packets to the different priority queues?). Our objective is to design such two mechanisms in a way that their overall behavior approximates the one of the PIFO queue. This is, an admission-control mechanism that (ideally) admits the earliest-arrived lowest-rank packets, and a queue-mapping algorithm that (ideally) prioritizes packets with lower rank.

\subsection{High-level intuition}
\label{ssec:intuition}

We introduce the high-level intuition behind \name's design by analyzing the case in which a PIFO queue schedules a batch of $A$ packets. We assume that all packets have the same size, and that the PIFO queue has a capacity of $B$ packets. For each incoming packet, the PIFO queue decides whether to admit the packet or drop it. Only after \emph{all} the packets have been processed, the PIFO queue schedules the admitted packets. 

\begin{figure}
	\centering
	\includegraphics[width=0.43\textwidth,keepaspectratio]{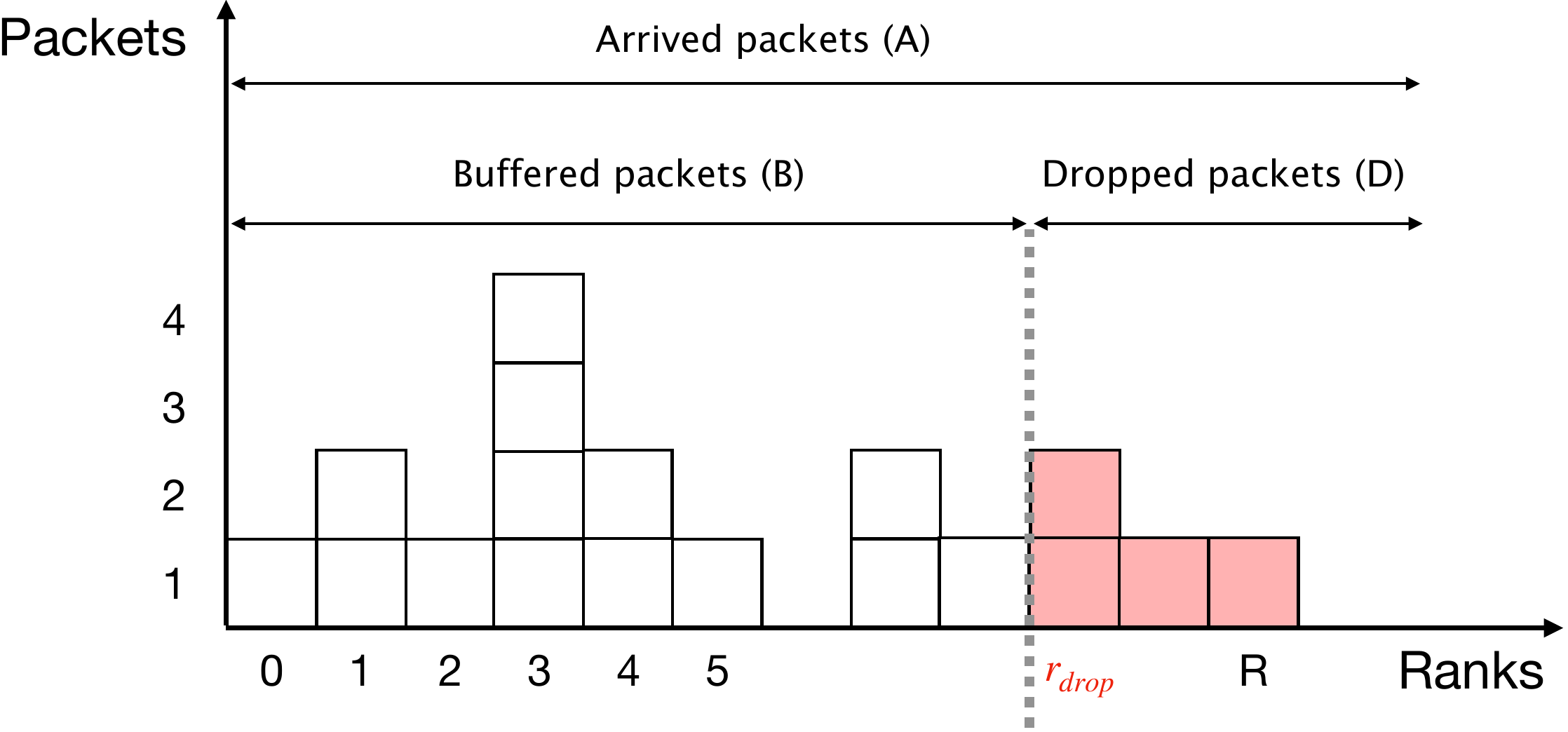}
	\caption{Admission control for a rank distribution, $\mathcal{W}$.}
	\label{fig:rdrop}
\end{figure}

\myitem{Approximating PIFO's admission} In this setup, the PIFO queue admits the $B$ (earliest-arrived) lowest-rank packets to the buffer, dropping the rest. Considering the rank distribution of the packets in the batch, $\mathcal{W}$, the admitted packets are the first $B$ packets that we find when reading the distribution from left to right (see Fig.~\ref{fig:rdrop}). As such, we can define a rank $r_{drop}$, such that all the packets with rank equal or higher than $r_{drop}$ are dropped by PIFO. Formally, computing $r_{drop}$ is finding the highest rank in the distribution, for which the quantile of the distribution stays below the available buffer capacity, $B$:
\begin{equation}
	\begin{split}
		\text{maximize}~~r_{drop},~~\text{where}~~0~\leq~r_{drop}~\leq~R,
		\\
		s.t.,~~\mathcal{W}.quantile(r_{drop} - 1) \leq B
	\end{split}
\end{equation}

Once $r_{drop}$ is known, approximating the PIFO's admission behavior on top of the \name abstraction is straightforward: we just need to configure its admission-control to drop all the packets from the batch with rank higher or equal than $r_{drop}$.

We note that, in some distributions, multiple packets may share the same rank. In that case, PIFO's admission control does not allow all packets with rank lower than $r_{drop}$, but only does so for the subset of them that have the smallest enqueue time. Thus, we can also define a time value, $t_{drop}$, above which PIFO drops all the packets of the highest-admitted rank (i.e., $r_{drop}-1$). We can approximate this behavior on the \name abstraction by configuring its admission-control to drop packets based on both, $r_{drop}$ and $t_{drop}$. Specifically, \name should drop packets if $r \geq r_{drop}$ or if $\{r=r_{drop}-1$ and $t \geq t_{drop}\}$.

\myitem{Approximating PIFO's scheduling} Once PIFO has decided whether to admit or drop each of the packets, it schedules the $B$ buffered packets in increasing order of rank and order of arrival (i.e., earliest-arrived lowest-rank first). This requires perfect sorting of packets at line rate. We can approximate this behavior on the \name abstraction by using strict-priority queues. When the number of queues is greater or equal than the number of ranks, we can perfectly sort packets by mapping each rank to a different priority queue. When that is not the case, we can still aim for a good approximation~\cite{sp-pifo}. 

\begin{figure}
	\centering
	\includegraphics[width=0.43\textwidth,keepaspectratio]{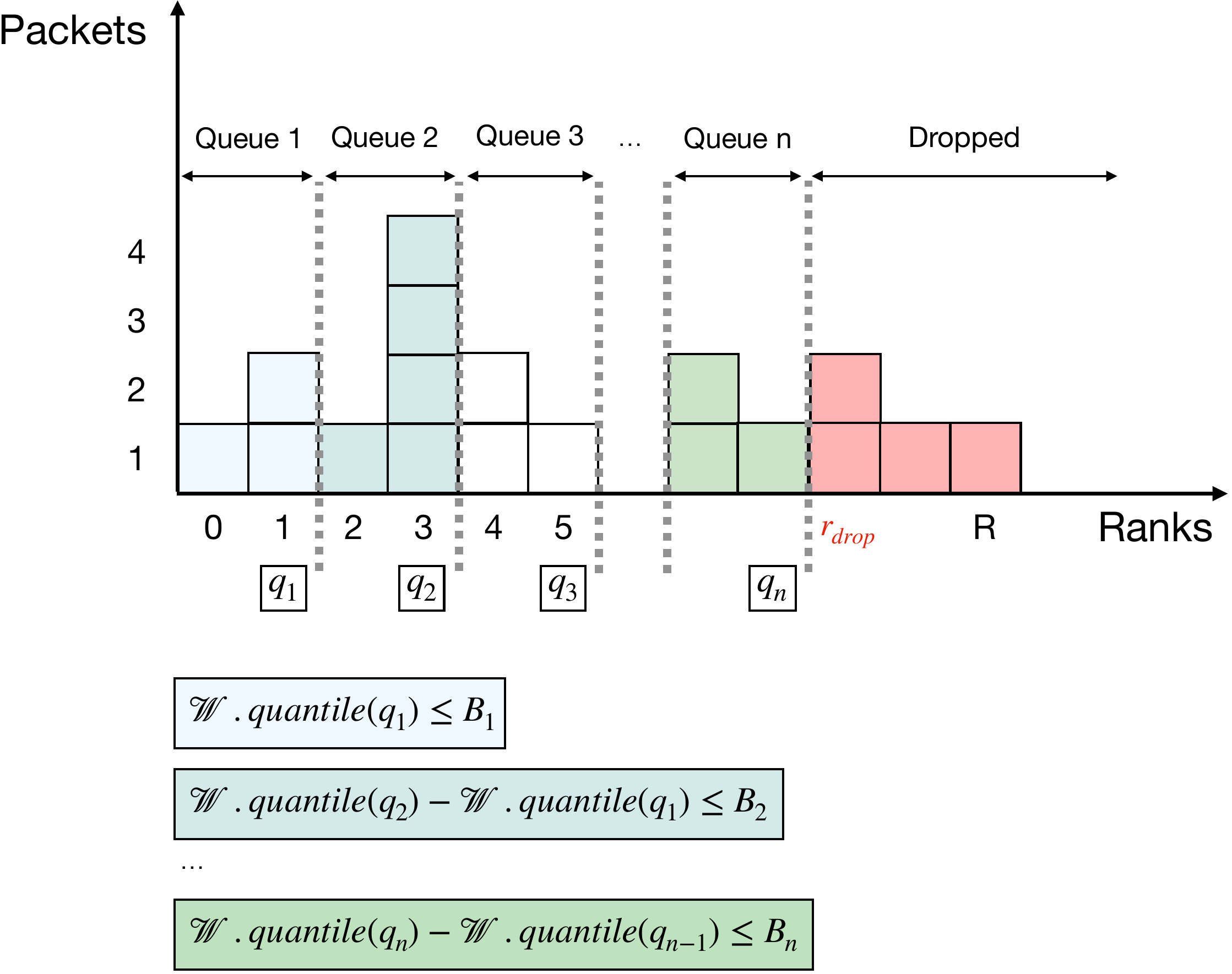}
	\caption{Queue mapping for a rank distribution, $\mathcal{W}$.}
	\label{fig:pieces}
\end{figure}

For each priority queue, $i$,  we define a rank, $q_{i}$, such that we only admit to the queue packets with rank lower or equal than $q_{i}$ (c.f., Fig.~\ref{fig:pieces}). We call these ranks \textit{queue bounds}. Formally, we let $\mlvec{q} =(q_1, \dots, q_n) \in \mathbb{Z}^n$ be the set of bounds for queues $1$ to $n$. We define a mapping strategy that uses queue bounds to map packets to their \emph{highest-possible} priority queue, based on their rank. For each incoming packet with rank $r$, we scan queues top-down (i.e., from highest- to lowest-priority) and map the packet to the first queue, $i$, that satisfies $r \leq q_{i}$.\footnote{Note that \name scans queues top-down, while SP-PIFO bottom-up~\cite{sp-pifo}.} With this definition, we convert the problem of sorting packets at line rate based on their ranks to the problem of finding the optimal queue bounds that maximize rank order at the output of the scheduler. We define a loss function $\mathcal{U_S}: \textrm{R}^n \times \textrm{R}_{\geq 0} \to \textrm{R}_{\geq 0}$, which stands for \emph{scheduling unpifoness}, such that $\mathcal{U_S}(\mlvec{q}, r)$ quantifies the approximation error of scheduling a packet with rank $r$ based on queue bounds $\mlvec{q}$ compared to an ideal PIFO queue. Intuitively, it measures the expected number of times that a packet with rank $r$ is scheduled after a packet with higher rank, $r'$. In the PIFO queue, $\mathcal{U_S} = 0$, since packets are scheduled in perfect order. Thus, in the \name abstraction, a lower $\mathcal{U_S}$ leads to a better approximation. 

Our goal is to find the optimal queue bounds, $\mlvec{q}_\mathcal{S}^{~*}$, that minimize $\mathcal{U_S}$. Let $\mathcal{Q}$ be the space of all valid queue-bound vectors and $\mathcal{W}$ the distribution of packet ranks. Then, $\mlvec{q}_\mathcal{S}^{~*}$ are:
\begin{equation}
	\mlvec{q}_\mathcal{S}^{~*} = \argmin_{\mlvec{q} \in \mathcal{Q}} \mathcal{U_S}(\mlvec{q}, r)
\end{equation}

Given that queue bounds are \emph{fixed} during the enqueue process, scheduling errors can not occur between ranks mapped to different priority queues. Thus, we can compute the total scheduling unpifoness as the sum of the individual losses at each priority queue. Letting $\mathcal{U}_\mathcal{S}(q_i)$ be the loss function corresponding to the queue with bound $q_i$, this is:
\begin{equation}
	\mathcal{U_S}(\mlvec{q}, r) = \sum_{1 \leq i \leq n}{\mathcal{U_S}(q_i)}
\end{equation}

Finally, letting $p_\mathcal{W}(r)$ and $p_\mathcal{W}(r')$ be the probability of ranks $r$ and $r'$, respectively, both mapped to the queue $i$, we can define the scheduling unpifoness of the queue as:
\begin{equation}
	\mathcal{U_S}(q_i) = \sum_{\substack{q_{i - 1} < r \leq q_{i}\\r < r' \leq q_{i}}}{p_\mathcal{W}(r) \cdot p_\mathcal{W}(r')}
\end{equation}

With this formulation, given that we know the exact rank distribution, $\mathcal{W}$, we can easily compute the optimal queue bounds, $\mlvec{q}_\mathcal{S}^{~*}$. For instance, \cite{gabor} proposes a modified version of the Bellman-Ford algorithm that does so in polynomial time.

To provide a high-level intuition about the optimal queue bounds, we derive an upper-bound of $\mathcal{U_S}(q_i)$ by setting $p_\mathcal{W}(r') = 1$. In doing so, we assume the worst case scenario in which, for each rank $r$, there is always a higher-rank packet, $r'$ in the queue that can produce a scheduling error. As such: 
\begin{equation}
	\begin{split}
		\hat{\mathcal{U_S}}(q_i) = \sum_{q_{i - 1} < r \leq q_{i}}{p_\mathcal{W}(r)} \\ = \mathcal{W}.quantile(q_i) - \mathcal{W}.quantile(q_{i-1})
	\end{split}
	\label{eq:queueboundsreordering}
\end{equation}

We can see how the optimal bounds are those that minimize the quantiles of the rank distribution for the set of ranks mapped to each priority queue. In other words, the optimal bounds are those that result in the least amount of different-rank packets mapped to each queue (i.e., those that minimize the colored area within each priority queue in Fig.~\ref{fig:pieces}).

Since we have to map all the admitted ranks, $0 \leq r < r_{drop}$, to some queue, removing a rank from a queue implies adding it to the adjacent queue. Thus, any reduction of unpifoness in a queue, increases the unpifoness of the adjacent queue. Therefore, we can only perform such an optimization step as long as there is a queue that can absorb the cost of taking in more ranks without becoming a new, greater maximum-cost queue. The optimum is achieved when the estimated scheduling unpifoness in each priority queue is balanced out. 

\myitem{\name's collateral drops} Unlike in PIFO, the admission-control in the \name abstraction (i.e., drop if $r \geq r_{drop}$) is not its only source of packet drops. Indeed, an admitted packet can still be dropped by the priority-queue enqueue mechanism if the available buffer space in the selected queue is not sufficient to accommodate the packet. As such, in order for the \name abstraction to fully approximate PIFO's admission behavior, it should not only control which packets are admitted; it should also make sure that the admitted packets are not dropped at enqueue when they are mapped to the priority queues. This brings us to the third part of the PIFO-approximation problem: approximating PIFO's efficient usage of the buffer space.

In the following, we compute the optimal queue bounds $\mlvec{q}_\mathcal{D}^{~*}$ that minimize the drops that occur when mapping packets to priority queues, and compare them to the optimal bounds that optimize rank order at the output of the scheduler, $\mlvec{q}_\mathcal{S}^{~*}$.

Let $B_i$ define the buffer capacity of the $i$-th priority queue in the \name abstraction. Let $\mlvec{B}=(B_1, \dots, B_n) \in \mathbb{Z}^n$ describe the buffer allocation across queues, where the sum of the buffer space of each queue is the total buffer space: $\sum_{i=1}^{n}{B_i} = B$. Let $\mlvec{q} = (q_1,...,q_n) \in \mathbb{Z}^n$ be the set of queue bounds defining the mapping strategy, where $0 \leq q_1 \leq q_2 \leq, \dots, \leq q_n = r_{drop} - 1$. With this strategy, we can compute the number of packets mapped to the $i$-th priority queue, $m_i$, as:
\begin{equation}
	\begin{split}
		m_1 = \mathcal{W}.quantile(q_1)
		\\
		m_2 = \mathcal{W}.quantile(q_2) - m_1
		\\
		m_n = \mathcal{W}.quantile(q_n) - m_{n-1}
	\end{split}
\end{equation}

We define a loss function $\mathcal{U_D}: \textrm{R}^n \times \textrm{R}_{\geq 0} \to \textrm{R}_{\geq 0}$, which stands for \emph{dropping unpifoness}, such that $\mathcal{U_D}(\mlvec{q})$ measures the number of packets dropped when mapping packets to queues based on queue bounds $\mlvec{q}$. In the PIFO queue, $\mathcal{U_D} = 0$, since there is no queue mapping, and drops only occur at admission. In \name, a lower $\mathcal{U_D}(\mlvec{q})$ leads to a better approximation. 

Our goal is to find the optimal bounds, $\mlvec{q}_\mathcal{D}^{~*}$, that minimize $\mathcal{U_D}(\mlvec{q})$. Let $\mathcal{Q}$ be the space of valid queue-bound vectors and $\mathcal{W}$ the distribution of packet ranks, then the bounds $\mlvec{q}_{\mathcal{D}}^{~*}$ are:
\begin{equation}
	\mlvec{q}_{\mathcal{D}}^{~*} = \argmin_{\mlvec{q} \in \mathcal{Q}} \mathcal{U_D}(\mlvec{q})
\end{equation}

Since queue bounds are fixed during the enqueue process, and packets are dropped in each queue independently of the other queues, we can compute the total unpifoness as the sum of the individual losses at each queue, $\mathcal{U_D}(q_i)$:
\begin{equation}
	\mathcal{U_D}(\mlvec{q}) = \sum_{1 \leq i \leq n}{\mathcal{U_D}(q_i)}
\end{equation}

The loss at queue $i$, $\mathcal{U_D}(q_i)$, is either the difference between the number of packets mapped to the queue, $m_i$, and the queue space, $B_i$, if the number of packets mapped to the queue is greater than the queue space, or $0$, otherwise:
\begin{equation}
	\mathcal{U_D}(q_i) = \begin{cases} 
		m_i - B_i &\text{if $m_i > B_i$}\\
		0 &\text{otherwise.}
	\end{cases}
\end{equation}

As such, the optimal bounds $\mlvec{q}_{\mathcal{D}}^{~*}$ are the ones that minimize the difference between the number of packets mapped to each queue and the buffer size of the queue. Since all packet drops contribute equally to the loss function, there may exist multiple queue-bound vectors, $\mlvec{q}_{\mathcal{D}}^{~*}$, that result in an optimal number of drops. In fact, any set of queue bounds is optimal as long as the packets mapped to each queue is lower or equal than the buffer space allocated to that queue (i.e., $m_i \leq B_i$):
\begin{equation}
	\forall i: \mathcal{W}.quantile(q_{i}) - \mathcal{W}.quantile(q_{i-1}) \leq B_i 
	\label{eq:queueboundsdrops}
\end{equation}

Given that the admission control in \name already makes sure that the total number of packets admitted can fit within the total buffer space (i.e., $\mathcal{W}.quantile(r_{drop}-1) \leq B$), we can guarantee that there exists at least one set of queue bounds, $\mlvec{q}_{\mathcal{D}}^{~*}$, that leads to zero drops at queue-mapping time. We can find such optimal bounds by computing the ranks for which the quantile of the rank distribution stays below the allocated queue sizes. This is $\forall i: \text{maximize}~q_{i}$ s.t. the eq.~\ref{eq:queueboundsdrops} is satisfied. 

Same as it happened in the admission-control counterpart, there may be rank distributions for which the number of packets of a certain rank exceeds the queue capacity (even when that rank is the only one mapped to the queue). In that case, we need finer granularity than the rank-level to perform the queue mapping. Same as we did for admission control, we can overcome this limitation by introducing an enqueue-time value $t_{i}$, for each priority queue, $i$, such that packets are only admitted to the queue if: $r \leq q_{i} - 1$ or if $\{r=q_{i}$ and $t \leq t_{i}\}$. Packets not admitted to the queue $i$ are carried over to the next queue, $i+1$, which has to account them as part of its quantile. 

\myitem{Sorting vs. dropping} Having computed the optimal queue bounds that best approximate PIFO in optimizing rank order at the output, $\mlvec{q}_{\mathcal{S}}^{~*}$, and in minimizing packet drops at queue-mapping, $\mlvec{q}_{\mathcal{D}}^{~*}$, we can see that they are not always the same. Indeed, $\mlvec{q}_{\mathcal{S}}^{~*}$ minimizes the quantiles of the rank distribution for the ranks mapped to each priority queue, and $\mlvec{q}_{\mathcal{D}}^{~*}$ minimizes the difference between these quantiles and their respective queue sizes. Thus, which queue bounds should we use?

In general, we could pick any of the two options based on e.g., which of the two behaviors we believe is more important. However, since our goal is to design a \emph{programmable} scheduler, we select the option that \emph{generalizes} the most. We realize that $\mlvec{q}_{\mathcal{D}}^{~*}$ are not only the best bounds for minimizing packet drops at queue-mapping time, but \emph{also} the optimal bounds for \emph{scheduling} in case the rank distribution is not known a priori (see eq.\ref{eq:queueboundsreordering} and eq.\ref{eq:queueboundsdrops}). Indeed, if the rank distribution of incoming packets is not known, the optimal queue mapping that minimizes rank reordering is the one that distributes packets across queues proportionally to the queue sizes. Thus, $\mlvec{q}_{\mathcal{D}}^{~*}$ can be seen as a worst-case bound for $\mlvec{q}_{\mathcal{S}}^{~*}$, leading to a good performance in both dimensions, as we show in \S\ref{sec:evaluation}. As a result, we leverage $\mlvec{q}_{\mathcal{D}}^{~*}$, as the queue bounds for our design.

\subsection{Online adaptation} 
\label{ssec:onlineadaptation}

So far, we have assumed a simplified scenario in which packets arrive to the scheduler in a \emph{batch}-basis, and where we know the complete rank distribution of the batch at enqueue. In practice, however, packets arrive in a \emph{stream} fashion, and the scheduler needs to perform the admission- and enqueue-decision \emph{per-packet, at line rate}. In the following, we present a set of techniques to translate our high-level intuition to a practical algorithm that we can easily deploy into hardware.

\myitem{Sliding window to monitor rank distribution} In the online setup we do not know the rank distribution of incoming packets in advance. Same as~\cite{gabor,aifo}, we solve this challenge by monitoring the rank distribution of recently-received packets using a sliding window, and using this distribution, $W$, as an estimate for the rank distribution of incoming packets, $\mathcal{W}$.

\myitem{Queue-occupancy to estimate congestion} In the online case, the buffer should absorb the short-term mismatches in traffic arrival and departure rates. Since measuring rates directly is hard, we monitor the buffer occupancy and use it as an estimate of the congestion level.~\footnote{This is a common approach in queue-management~\cite{red, codel, aifo}. We could also have used the \emph{sojourn-time} of packets, as proposed by CoDel~\cite{codel}.} Given $b$, the buffer-occupancy level at a certain packet's enqueue time, we decide to enqueue the packet if $W.quantile(r_{drop} - 1) \leq \left[~\frac{1}{1-k} \cdot \frac{B-b}{B}~\right]$, where $k$ is a parameter that we can optionally use to allow for some burstiness. By applying the same technique to the queue-mapping algorithm, we can define the queue bounds, $\mlvec{q}$, as:
\begin{equation}
	\begin{split}
		W.quantile(q_{1}) \leq \left[~\frac{1}{1-k} \cdot \frac{B-b}{B} \cdot \frac{B_1}{B}~\right] \\
		W.quantile(q_{2}) \leq \left[~\frac{1}{1-k} \cdot \frac{B-b}{B} \cdot \frac{B_1 + B_2}{B}~\right] \\
		\cdots\\
		W.quantile(q_{n}) \leq \left[~\frac{1}{1-k} \cdot \frac{B-b}{B} \cdot \frac{\sum_{i=1}^{n}B_i}{B}~\right]
	\end{split}
\end{equation}

When queues have equal size (i.e., $B_i = B/n$),  the mapping condition for queue $i$ is: $~W.quantile(r) \leq \left[~\frac{1}{1-k} \cdot \frac{B-b}{B} \cdot \frac{i}{n}~\right]$. Note that, since $q_n = r_{drop} - 1$,  the queue-mapping policy for the lowest-priority queue is equivalent to the overall admission control at the scheduler. As such, the queue-mapping mechanism already implies the admission control, which simplifies the algorithm design and its implementation (\S\ref{ssec:algorithm}, \S\ref{sec:implementation}).

\myitem{Minimizing collateral drops} Same as in the batch case, there may be cases in which not all the packets of a certain rank can fit into the available space of a queue. In the batch case, we leveraged the notion of $t_{i}$ to enqueue packets of a certain rank to a lower-priority queue if the first queue where the rank could be admitted was already full. In the online case, accurately computing $t_{i}$ is not straightforward. We overcome this limitation by using queue-occupancy information at enqueue. Specifically, we modify the mapping process such that, if  the highest-priority queue at which a given packet should be mapped is already full, we enqueue the packet to the next queue that has available space. Since we scan queues top-down, we can guarantee that this action will preserve packet order within the same rank. Overall, this technique avoids dropping a low-rank packet that would be admitted by PIFO, at the cost of (at most) one scheduling inversion at the output. 

\subsection{\name algorithm}
\label{ssec:algorithm}

\begin{algorithm}[h]
	\caption{\name} \label{alg:per_packet}
	\begin{algorithmic}[1]
		\Require An incoming packet $pkt$ with rank $r$
		\Procedure{Ingress}{}
		\State Update sliding window $W$ with $r$
		\State $B \gets$ buffer.total
		\State $b \gets$ buffer.used
		\For{$Queues(i)$ : $i = 1$ \textbf{to} $n$} \Comment Scan top-down
		\If{$~W.quantile(r) \leq \left[~\frac{1}{1-k} \cdot \frac{B-b}{B} \cdot \frac{\sum_{i=1}^{n}B_i}{B}~\right]$}
		\If{~$Queues(i).notFull()$}
		\State $Queues(i).enqueue(pkt)$ \Comment Select queue
		\State \textbf{return;}
		\EndIf
		\EndIf
		\EndFor
		\State $Drop(pkt)$ \Comment Drop packet
		\EndProcedure
	\end{algorithmic}
	\label{alg:packs}
\end{algorithm}

We detail the \name algorithm in alg~\ref{alg:packs}. First, \name uses a sliding window to monitor the rank distribution of the latest packets enqueued. Second, it computes quantiles on the distribution to decide whether to admit incoming packets, and how to map admitted packets to the various priority queues. 

Whenever an incoming packet arrives, \name updates the sliding window, $W$, with the packet rank. Then, it measures the current buffer occupancy, $b$ and uses it to compute the portion of overall buffer space, $B$, that is still free: $\frac{B-b}{B}$. \name admits the incoming packet if the quantile of its rank is lower than the percentage of available buffer: $W.quantile(r) \leq \left[~\frac{1}{1-k} \cdot \frac{B-b}{B}~\right]$. Note that we weight the admission condition by an optional parameter $k$, to allow for some burstiness. Also note that, in alg.~\ref{alg:packs}, the admission condition is \emph{implicit} in the queue-mapping process. Indeed, the drop action in line $10$, executed when the packet has \emph{not} been mapped to the lowest-priority queue, does the same job as an explicit admission control. 

For the admitted packets, \name scans priority queues top-down (i.e., from highest- to lowest-priority) and maps the packet to the first queue that has available space and that satisfies the condition: $W.quantile(r) \leq \left[~\frac{1}{1-k} \cdot \frac{B-b}{B} \cdot \frac{\sum_{i=1}^{n}B_i}{B}~\right]$. If a packet is not admitted to any of the queues, because its rank is too-high, or because all queues are full, it is dropped.

In Appendix~\ref{sec:theoretical analysis}, we prove that \name is optimal under certain conditions pertaining to window and buffer sizes. To do so, we prove that the departure rate for all packets ranks is the same in \name as in a PIFO queue.
\section{Implementation}
\label{sec:implementation}

In this section, we describe our implementation of \name in \pfour for the Intel Tofino 2 Native Architecture. Our implementation follows the algorithm described in \S\ref{ssec:algorithm} and spans 529 lines of code. It uses only 12 out of the 20 available stages of Tofino 2. Table.~\ref{table:resources} summarizes the resource utilization. 

When a packet arrives, \name performs four operations: (i) it monitors the rank distribution of latest-arrived packets; (ii) it computes the quantiles of the distribution for the packet rank; (iii) it measures the available buffer size; and (iv) it checks the admission and queue-mapping conditions to decide whether to admit the packet or drop it, and to which queue to map it. 

\myitem{Monitoring the rank distribution} We monitor the rank distribution of latest-enqueued packets by implementing a sliding window over a set of $|W|$ registers. Each register stores the rank of one packet, and we use a circular packet counter, from $0$ to $|W|-1$, to track the position of the latest rank added to the window. For each arriving packet, we check the value of the counter, and over-write the register indicated by the counter with the value of the new packet rank. We then update the circular counter, resetting it to $0$ if it reaches the value of $|W|$. In our prototype, we use a sliding window of size $16$ (it could be increased by using sampling~\cite{aifo}), which uses $4$ stages and accesses $4$ registers in parallel at each stage.  

\myitem{Computing quantiles} For each incoming packet with rank $r$, we compute the quantile of the monitored rank distribution for the packet's rank, $r$, by counting the number of times $r$ is lower than a rank in the sliding window, and dividing the result by the total window size,  $|W|$. Specifically, whenever we access each register to monitor the sliding window, we also compare if the rank of the incoming packet is lower than the register value. We perform the comparison at the same register access as the sliding-window update, within the stateful ALU. We write the result of each comparison in a metadata field as a binary result, where for register $j$, $output_j = 1$ if $r<value_j$, and $0$, otherwise. We then compute the distribution quantile for the rank, by adding up all the output values, and dividing the result by the total number of registers,$|W|$:  $W.quantile(r)=(\sum_i{output_i})/|W|$. We aggregate all metadata fields by summing two results at each stage using non-stateful ALUs. This takes $log_2|W|$ stages, being $|W|$ the window size. We parallelize the last register accesses, with the first metadata sums to reduce the number of stages.

\myitem{Measuring the available buffer} Differently from previous works, we measure buffer occupancy levels at enqueue by leveraging a \emph{Ghost thread}, available in Tofino 2~\cite{queueawaretofino}. In the first generation of programmable switches, queue-occupancy information was only accessible from the egress pipeline, since packets were required to cross the traffic manager to obtain such information. As such, previous works were forced to engineer packet-recirculation techniques to make this information available at the ingress pipeline, at the cost of reducing throughput and processing resources~\cite{aifo}. Instead, we use a Ghost thread to make the queue-utilization information at the ingress: we define a Ghost control that periodically writes the queue depth provided by the traffic manager to a register, and we directly access this information from the ingress pipeline. 

\myitem{Deciding admission and queue-mapping} After computing the quantile of the rank distribution for the incoming packet's rank, $W.quantile(r)$, and having measured the available buffer, $b$, we combine them to decide whether to admit the packet or drop it, and how to map it to priority queues. In our prototype, we leverage four priority queues and assign the same buffer size to each queue, such that the admission and mapping conditions are easier to compute. For each queue $i$, we need to compute: $W.quantile(r) \leq \frac{1}{1-k} \cdot \frac{B-b}{B} \cdot \frac{i}{n}$ (cf.~\S\ref{sec:design1}). We convert the equation into $W.quantile(r) \cdot (1-k) \cdot B \cdot n \leq (B-b) \cdot i$. Then, we compute the right side of the equation by using the math unit and bit-shift operations. For the four queues, this takes three stages. We compute the left side of the equation by picking the value of $k$ strategically, such that the whole operation can be performed by a bit shift. Finally, we compute the \textit{minimum} operation for each of the queues, and execute the corresponding drop or enqueue action based on its result. 

\begin{table}
	\centering
	\begin{tabularx}{0.45\textwidth}{lX} \toprule
		Resource Type & Usage (Average per stage) \\ \midrule
		Exact Match Crossbar & 3.4~\% \\
		Gateway & 3.4~\% \\
		Hash Bit & 1.3~\% \\
		Hash Dist. Unit & 4.2~\% \\
		Logical Table ID & 10.9~\% \\
		SRAM & 2.4~\% \\
		TCAM & 0~\% \\
		Stateful ALU & 23.8~\% \\
		\bottomrule
	\end{tabularx}
	\caption{Resource requirements of \name on Intel Tofino 2.}
	\label{table:resources}
\end{table}
\begin{figure*}
	\centering
    \begin{subfigure}{0.2356\textwidth}
		\centering
		\includegraphics[width=\textwidth,keepaspectratio]{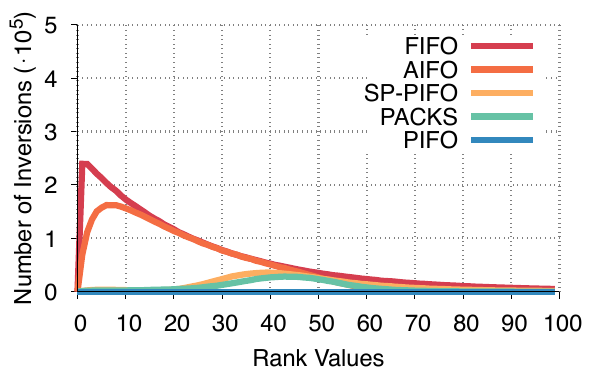}
		\subcaption{Exponential (inversions)}
		\label{fig:theoretical_a}
	\end{subfigure}
    \begin{subfigure}{0.2356\textwidth}
		\centering
		\includegraphics[width=\textwidth,keepaspectratio]{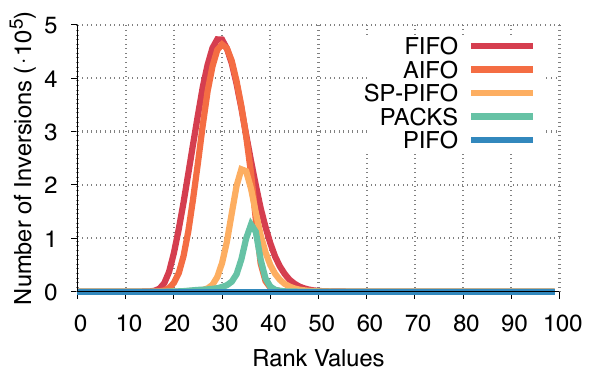}
		\subcaption{Poisson (inversions)}
		\label{fig:theoretical_b}
	\end{subfigure}
	\begin{subfigure}{0.2356\textwidth}
		\centering
		\includegraphics[width=\textwidth,keepaspectratio]{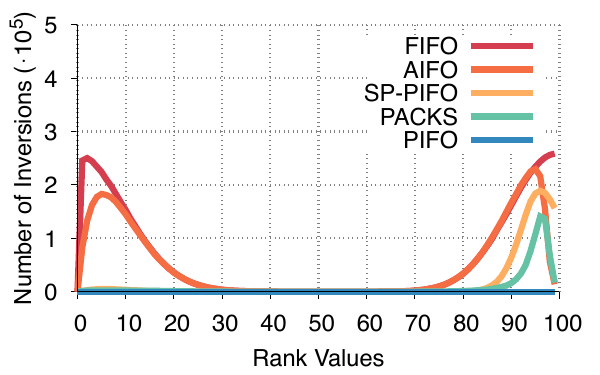}
		\subcaption{Convex (inversions)}
		\label{fig:theoretical_c}
	\end{subfigure}
	\begin{subfigure}{0.2356\textwidth}
		\centering
		\includegraphics[width=\textwidth,keepaspectratio]{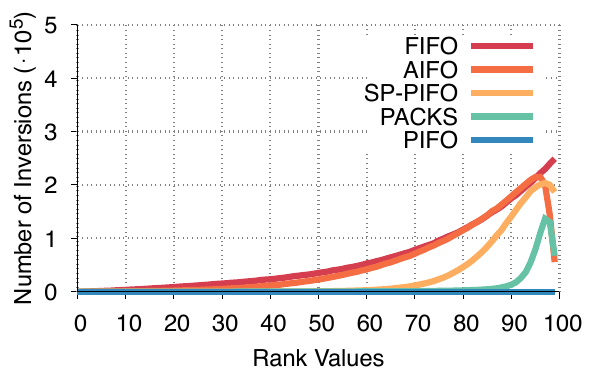}
		\subcaption{Inv. Exponential (inversions)}
		\label{fig:theoretical_d}
	\end{subfigure}
	
	\begin{subfigure}{0.2356\textwidth}
		\centering
		\includegraphics[width=\textwidth,keepaspectratio]{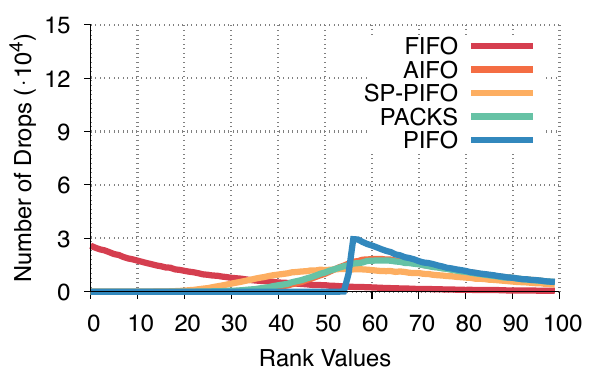}
		\subcaption{Exponential (drops)}
		\label{fig:theoretical_e}
	\end{subfigure}
    \begin{subfigure}{0.2356\textwidth}
		\centering
		\includegraphics[width=\textwidth,keepaspectratio]{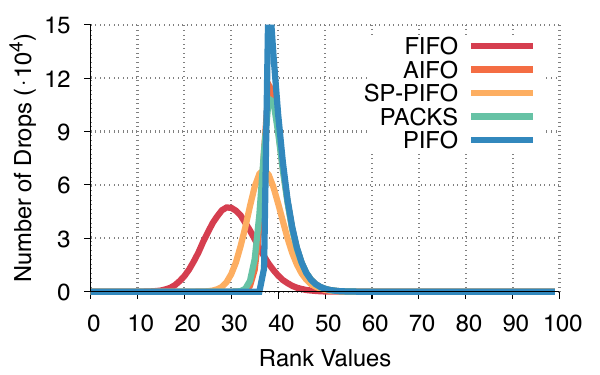}
		\subcaption{Poisson (drops)}
		\label{fig:theoretical_f}
	\end{subfigure}
	\begin{subfigure}{0.2356\textwidth}
		\centering
		\includegraphics[width=\textwidth,keepaspectratio]{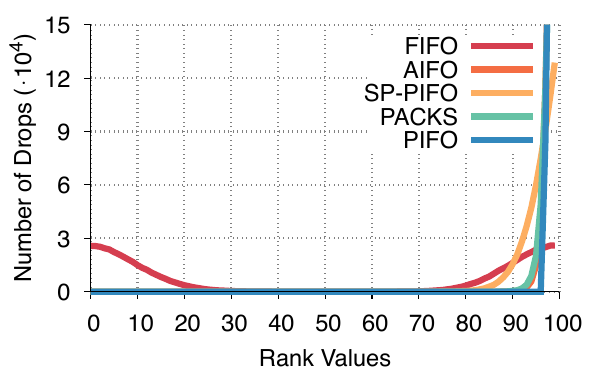}
		\subcaption{Convex (drops)}
		\label{fig:theoretical_g}
	\end{subfigure}
	\begin{subfigure}{0.2356\textwidth}
		\centering
		\includegraphics[width=\textwidth,keepaspectratio]{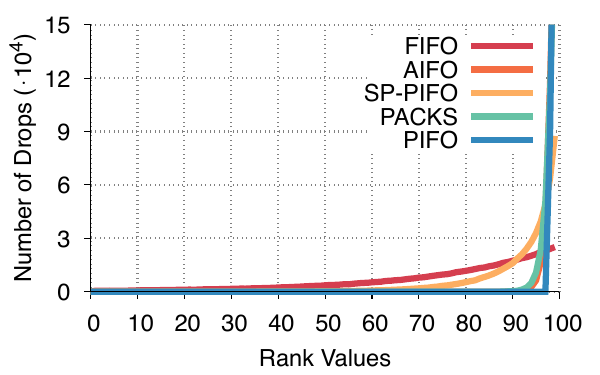}
		\subcaption{Inv. Exponential (drops)}
		\label{fig:theoretical_h}
	\end{subfigure}
	\caption{Rank inversions and packet drops when scheduling alternative rank distributions (8 queues).}
	\label{fig:theoretical}
\end{figure*}

\section{Evaluation}
\label{sec:evaluation}

We now evaluate \name performance and practicality. We first use packet-level simulations to evaluate how \name approximates PIFO's \emph{scheduling} and \emph{admission} behaviors for various rank distributions (\S\ref{ssec:evaluation:theoreticalanalysis}). Second, we show how \name approximates scheduling objectives under realistic traffic workloads (\S\ref{sec:evaluation:performance}). Finally, we evaluate \name performance when deployed on hardware switches (\S\ref{sec:evaluation:hardware}).

\subsection{Behavioral analysis}
\label{ssec:evaluation:theoreticalanalysis} 

First, we evaluate \name's performance in approximating PIFO behaviors for different rank distributions. We compare its behavior to that of an ideal PIFO queue, along with other scheduling schemes for reference:  FIFO,  SP-PIFO, and AIFO.

\myitem{Methodology} We implement the schedulers in Netbench~\cite{netbench}, a packet-level simulator. We analyze the performance of a single switch, when it schedules a constant bit-rate flow of 11~Gbps over a bottleneck link of 10~Gbps, during one second. We tag each packet with a rank within  $[0-100]$, drawn from either an exponential, Poisson, convex, or inverse-exponential distribution. We set \name and SP-PIFO to have 8 priority queues, each with 10 packets of capacity. We configure AIFO and FIFO with a single queue with capacity for 80 packets. We configure \name and AIFO with a window size, $|W|$, of 20 packets, and a burstiness allowance, $k$, of 0.

We measure the number of scheduling inversions generated by each rank (i.e., the number of times a packet of a certain rank is scheduled \emph{before} a packet with lower rank in the queue), and the number of packets dropped of each rank. 

\myitem{Uniform case} In \S\ref{ssec:limitations} we have seen how \name outperforms existing schemes in the case of a uniform rank distribution. On the one hand, by leveraging multiple priority queues, \name improves the rank order at the output with respect to single-queue schemes. It also outperforms the state-of-the-art multi-queue scheme (i.e., SP-PIFO) by estimating the rank distribution and using it to best map packets to priority queues. On the other hand, by proactively dropping high-rank packets, \name also achieves the closest-to-PIFO drop distribution. 

We now extend the study to other rank distributions. Fig.~\ref{fig:theoretical}  depicts the scheduling inversions and packet drops generated by each scheduler across ranks for the various rank distributions. We observe how, for all the rank distributions, \name outperforms existing schemes (SP-PIFO and AIFO), and gets closest to PIFO in scheduling inversions and packet drops.

\myitem{Minimizing inversions} Single-queue schemes (i.e., FIFO and AIFO) perform poorly in terms of scheduling inversions across all ranks for all distributions, due to their inability to prioritize traffic. 
AIFO performs slightly better than FIFO, reducing the number of inversions by $22\%$, $21\%$, $18\%$, and $15\%$ for the exponential, Poisson, convex and inverse-exponential distributions, respectively. SP-PIFO further reduces inversions by mapping low-rank packets to higher-priority queues. With respect to AIFO, SP-PIFO reduces the number of inversions by up to $68\%$ (convex) and $67\%$ (Poisson). \name significantly outperforms the other schemes by further reducing the number of scheduling inversions. With respect to SP-PIFO, \name still manages to reduce the number of inversions by $33\%$, $64\%$, $54\%$, and $75\%$, for the exponential, Poisson, convex, and inverse-exponential distributions, respectively.

\myitem{Approximating the drop distribution} \name also outperforms in approximating PIFO's \emph{distribution} of packet drops across ranks (i.e., the fact that it always drops the highest-rank packets in the distribution). \name and AIFO consistently approximate this behavior across all distributions, while SP-PIFO tends to also drop packets with lower-ranks. For example, for the inverse-exponential distribution, the lowest-rank packet dropped by PIFO, \name, AIFO, SP-PIFO, and FIFO has a rank of 98, 80, 66, 16, and 1 respectively. We observe a similar behavior for all the distributions. The lowest rank dropped by \name \emph{overall} is rank $30$ in the Poisson distribution. Even in that case, \name is still the closest to PIFO. PIFO's lowest rank dropped is 37, and the lowest rank dropped for AIFO, SP-PIFO and FIFO are 29, 21 and 7, respectively.

\subsection{Performance analysis}
\label{sec:evaluation:performance}

We extend the packet-level simulations to evaluate \name's performance under two practical scenarios. We consider two scheduling objectives: (i) minimizing flow completion times; and (ii) enforcing fairness. In both cases, we show that \name achieves near-optimal performance.

\myitem{Methodology} Same as previous work~\cite{pfabric,sp-pifo, aifo, hcsfq}, we use a leaf-spine topology with 144 servers connected through 9 leaf and 4 spine switches, and set the access and leaf-spine links to 1Gbps and 4Gbps, respectively. We generate traffic flows following the pFabric web-search workload~\cite{pfabric}. Flow arrivals are Poisson-distributed and we adapt their starting rates to achieve different utilization levels. We use ECMP and draw source-destination pairs uniformly at random.

\begin{figure*}
	\begin{subfigure}{0.33\linewidth}
		\centering
		\includegraphics[width=\linewidth,keepaspectratio]{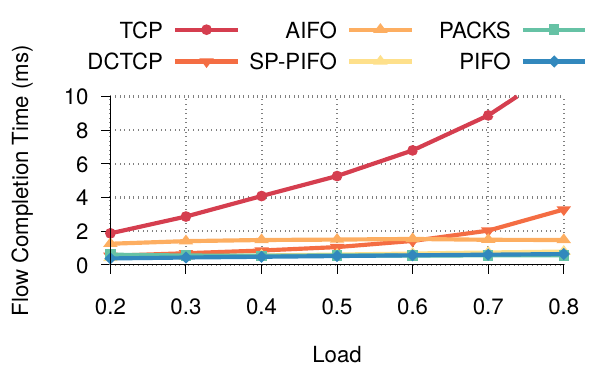}
		\subcaption{(0,100KB): Average FCT}
		\label{subfig:pFabric_a}
	\end{subfigure}\hfill
	\begin{subfigure}{0.33\linewidth}
		\centering
		\includegraphics[width=\linewidth,keepaspectratio]{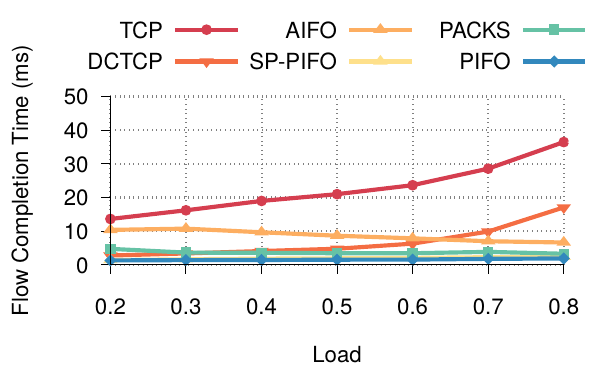}
		\subcaption{(0,100KB): 99th percentile FCT}
		\label{subfig:pFabric_b}
	\end{subfigure}\hfill
	\begin{subfigure}{0.33\linewidth}
		\centering
		\includegraphics[width=\linewidth,keepaspectratio]{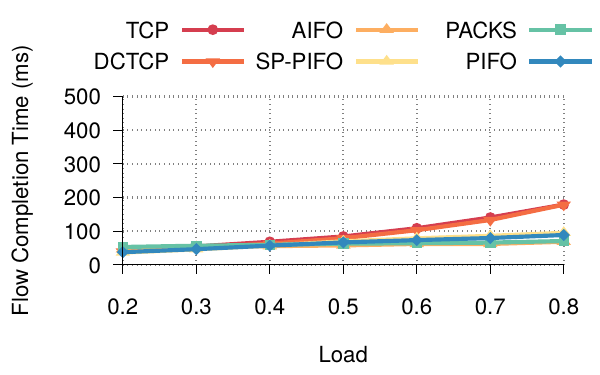}
		\subcaption{[1MB,$\infty$): Average FCT}
		\label{subfig:pFabric_c}
	\end{subfigure}
	
	\caption{pFabric: FCT statistics across different flow sizes in web search workload.}
	\label{fig:pFabric}
\end{figure*}
\begin{figure*}
	\begin{subfigure}{0.33\textwidth}
		\centering
		\includegraphics[width=\textwidth,keepaspectratio]{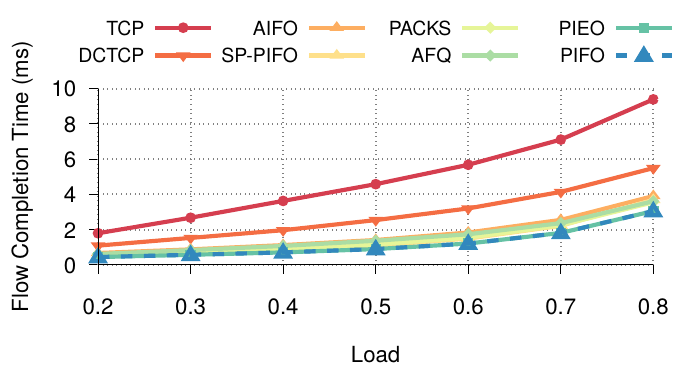}
		\subcaption{(0,100KB): Average FCT}
		\label{subfig:fairness_a}
	\end{subfigure}
	\begin{subfigure}{0.33\textwidth}
		\centering
		\includegraphics[width=\textwidth,keepaspectratio]{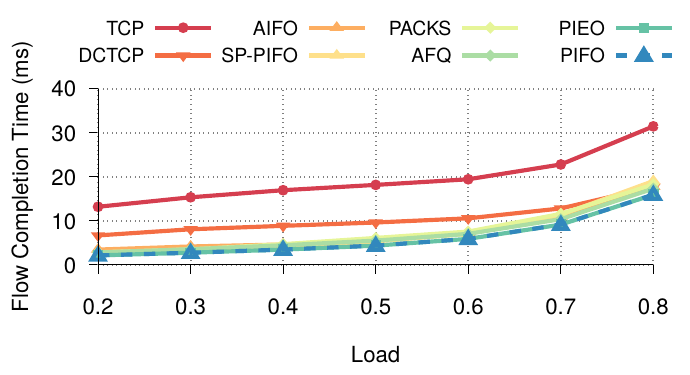}
		\subcaption{(0,100KB): 99th percentile FCT}
		\label{subfig:fairness_b}
	\end{subfigure}
	\begin{subfigure}{0.33\textwidth}
		\centering
		\includegraphics[width=\textwidth,keepaspectratio]{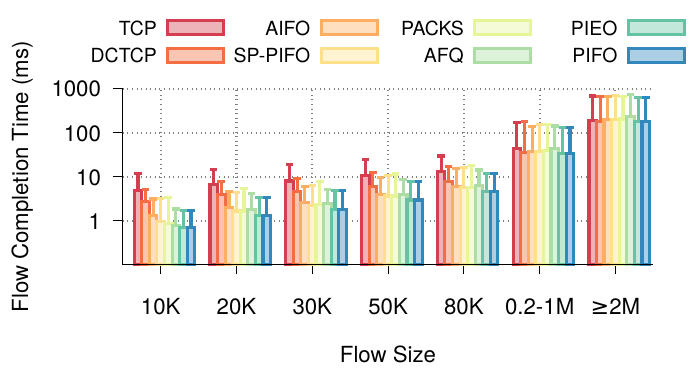}
		\subcaption{FCT breakdown for 70\% load}
		\label{subfig:fairness_c}
	\end{subfigure}	
	\caption{Fairness: FCT statistics for all flows at different loads, over the web search workload.}
	\label{fig:fairness1}
\end{figure*}

\subsubsection{Minimizing Flow Completion Times}
\label{sssec:fct}
\myitem{Rank definition \& benchmarks} We run pFabric~\cite{pfabric} on top of PIFO, AIFO, SP-PIFO, and \name, and evaluate their performance in minimizing flow completion time. pFabric sets packet ranks based on their remaining flow sizes. As recommended in~\cite{pfabric}, we approximate pFabric's rate control through standard TCP with a retransmission timeout of 3 Round Trip Times (RTTs), and equalize the difference in RTOs between schemes with the proportional queue size. Indeed, we utilize an RTO of 96$\mu$s and 8 queues$\times$10 packets for \name and \textsf{SP-PIFO} (respectively, 1 queue$\times$80 packets in PIFO and AIFO), and an RTO of 300$\mu$s and 146KB drop-tail queues for TCP and DCTCP, with ECN marking threshold set at 14.6KB. For \name and AIFO, we set the window size, $|W|$ to 80 packets, and the burstiness-allowance, $k$, to 0.

\myitem{Results} Fig.~\ref{fig:pFabric} depicts the average and 99th percentile FCTs of large ($\geq 1$MB) and small flows ($<100$KB). We see that TCP results in high FCTs for small flows, both in average ($1.88$ to $11.82~ms$) and at the 99th percentile ($13.63$ to $36.45~ms$) due to not prioritizing packets based on their ranks. DCTCP reduces such FCTs by leveraging ECN-marking up to $0.53$--$3.29~ms$ (average) and $2.77$--$17.05~ms$ (99-th percentile). At very low-loads AIFO's proactive dropping of packets is detrimental for performance, even worse than DCTCP for small flows. However, AIFO improves the performance with respect to DCTCP for high loads (e.g., decreases the average FCT of small flows with respect to DCTCP by $\sim50\%$ at 80\% utilization). Still, due to not being able to prioritize packets based on ranks AIFO's performance is still far from optimal ($\sim3\times$ higher average FCTs for small flows than PIFO). 

By efficiently prioritizing low-rank packets, both SP-PIFO and \name achieve consistently good performance across flow sizes. While they achieve very similar performance, \name outperforms in terms of average FCTs for higher loads, while SP-PIFO adapts better under lower loads. SP-PIFO stays between $7.35$ and $23.14\%$ of average FCTs for small flows difference with respect to PIFO. Under low load, SP-PIFO outperforms \name (e.g., 17.2\% better than PACKS at 80\% load in the 99th percentile), since the expected distribution may not match the real one, and \name's proactive dropping (especially under the no-bursts allowed policy, $k=0$) impacts performance. SP-PIFO benefits diminish under very high loads (e.g., $23.14\%$ on average FCT and $24.74\%$ in the 99-th percentile with respect to PIFO at 80\% utilization). Under high load, \name outperforms SP-PIFO, being only $11.77\%$ away from PIFO, closer than the $23.14\%$ of SP-PIFO.

\subsubsection{Enforcing fairness across flows}

\myitem{Rank definition \& benchmarks} We now evaluate how the various schedulers perform at enforcing fairness across flows. To that end, we run the Start-Time Fair Queueing (STFQ) rank design~\cite{stfq} on top of PIFO, \name, SP-PIFO, and AIFO. We compare their performance to the one of AFQ~\cite{afq} and PIEO~\cite{pieo} for reference (\S\ref{sec:relatedwork}). We now use 32 queues$\times$10 packets in SP-schemes (resp.
1 queue$\times$320 packets for single-queue schemes). We set the bytes-per-round of AFQ to 80 packets. Same as previous works~\cite{sp-pifo,aifo}, we generate traffic following the pFabric web-search distribution, and evaluate fairness by measuring the flow completion time of short flows. 

\myitem{Summary} Fig.~\ref{subfig:fairness_a} and Fig.~\ref{subfig:fairness_b}
depict the average and the 99-th percentile of the flow completion times for small flows across different levels of utilization (20\% to 80\%). Fig.~\ref{subfig:fairness_c} depicts the FCTs across flow sizes at 70\% utilization. In all cases, \name achieves near-PIFO behavior and is on-par with the state-of-the-art approaches (AFQ, SP-PIFO, and AIFO).

\paragraph{Impact of the utilization (Fig.~\ref{subfig:fairness_a} \& Fig.~\ref{subfig:fairness_b})} \name stays within $\sim$14--23\% in average FCT of the ideal PIFO, across all levels of utilization. In terms of average FCT for short flows, \name consistently outperforms TCP, DCTCP, AIFO and AFQ across all loads. Specifically, \name reduces the average FCTs for short flows by  62--75\%, 35--54\%, 9--21\%, 2--16\%  with respect to TCP, DCTCP, AIFO and AFQ, respectively. With respect to SP-PIFO, \name performs worse for lower loads, with differences between $\sim$0--6\%, but outperforms by 4\% for the highest load (80\% utilization).

At the 99th percentile, \name stays within $\sim$12--28\% of the ideal PIFO performance across all levels of utilization, performing very similarly to AIFO, SP-PIFO and AFQ (e.g., at 80\% load, these schemes perform within 5\% of each other). In that case, AFQ performs best, followed by AIFO and \name, and finally, SP-PIFO. At 80\% load, \name's performance is 4\% better than SP-PIFO, just 4.12\% away of the ideal. 

\myitem{Impact of flow sizes (Fig.~\ref{subfig:fairness_c})} The performance of \name at 70\% utilization stays
within $\sim$12--21\% of the ideal PIFO performance across all flow sizes, and is on-par with the other state-of-the-art approaches. Significantly, for the smallest flows ($<$10K), \name achieves the lowest average FCT (i.e., closest to PIFO), only outperformed by AFQ by a 10\%. 

\begin{figure}
	\centering
    \begin{subfigure}{0.2356\textwidth}
		\centering
		\includegraphics[width=\textwidth,keepaspectratio]{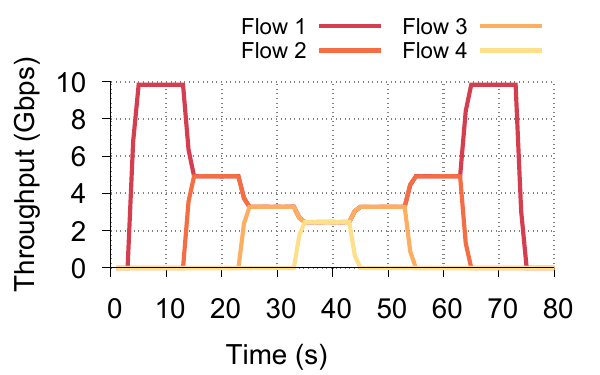}
		\subcaption{FIFO Bandwidth-split}
	\end{subfigure}
	\begin{subfigure}{0.2356\textwidth}
		\centering
		\includegraphics[width=\textwidth,keepaspectratio]{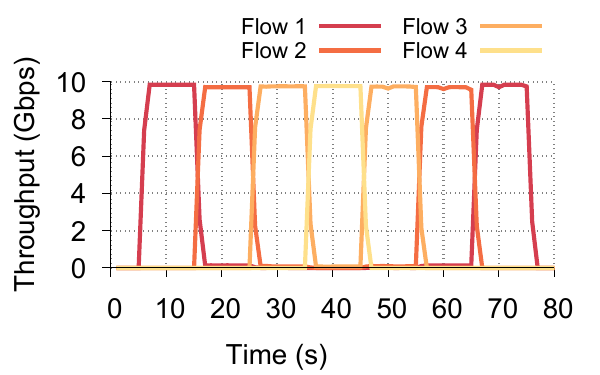}
		\subcaption{\name Bandwidth-split}
	\end{subfigure}
	\caption{Hardware testbed: Bandwidth allocation under progressive flow generation with increasing priorities.}
	\label{fig:tofino}
\end{figure}

\subsection{Hardware testbed}
\label{sec:evaluation:hardware}

We finally evaluate our hardware-based implementation of \name on the Edgecore Networks DCS810 (AS9516-32D) P4 Programmable Intel Tofino2 Switch~\cite{tofino}. Same as previous works~\cite{sp-pifo,aifo,hcsfq}, we evaluate the bandwidth allocated by \name to flows with different ranks when scheduled over a bottleneck link. We generate traffic between two servers, connected by a Tofino2 switch, using interfaces of 100~Gbps (sender$\rightarrow$switch) and 10~Gbps (switch$\rightarrow$receiver). We run four UDP flows of 20~Gbps each using 
MoonGen~\cite{moongen, moongen2}. We start the flows sequentially, in increasing order of priority (decreasing rank). We start a new flow every $10$ seconds, until all of them are running. Then, we wait for 10 seconds and start stopping the flows. We stop flows sequentially, in decreasing order of priority (i.e., starting by the one with highest-priority). We stop one flow every $10$ seconds, until stopping all flows.

Fig.~\ref{fig:tofino} depicts the flows' bandwidth and how \name manages to effectively prioritize traffic from lower ranks. While the FIFO queue distributes the bandwidth uniformly across flows (failing at prioritizing traffic), \name successfully allocates the available bandwidth to the highest-priority flow. 
\section{Related work}
\label{sec:relatedwork}

\myitem{Programmable packet scheduling} While packet scheduling has been extensively studied in the last decades~\cite{fq,sfq,wfqfifoplus,drr,stfq,pfabric,srpt,pias,afq}, the idea of making it programmable is relatively new~\cite{pifo0,ups}. \cite{pifo0} introduced the concept of programmable packet scheduling as a means to deploy any scheduling algorithm into hardware without need for new ASIC designs. At the same time, \cite{pifo0,pifo1} introduced the PIFO queue as the abstraction for programmable scheduling. While promising, implementing PIFO queues in hardware resulted difficult. As such, various proposals went on to approximate PIFO queues in existing programmable data planes: SP-PIFO~\cite{sp-pifo}, QCluster~\cite{qcluster}, PCQ~\cite{pcq}, AIFO~\cite{aifo}, Spring~\cite{gabor}, and Gearbox~\cite{gearbox}. \name builds on the general ideas behind these approaches (i.e., leveraging (multiple) FIFO queues with different priority, and deciding what packets to admit and how to map them to priority queues) and solves their limitations by proposing a unifying abstraction that outperforms at approximating PIFO. 

\myitem{Active queue management (AQM)} AQM schemes have also received attention for decades~~\cite{rfc7567,red,rfc8289,codel,rfc8033,rfc8290}. Recently, PR-AQM~\cite{praqm} has been proposed, as an AQM scheme that runs on top of a multi-queue programmable scheduler to control the delay of the different queues (\textit{à la CoDel}~\cite{codel}) while preserving the rank-based differentiated service. Even though programmable packet schedulers (e.g., AIFO and \name) and AQM schemes (e.g., PR-AQM) use similar techniques (e.g., rank-based admission control), their goal is different. Schedulers determine how to distribute the output capacity across packets, by deciding what packet to send next and when to do so. Instead, AQM schemes control the sizes of queues that build in network buffers and minimize the time that packets spend in them. While complementary, they tackle different problems and work at different time scales~\cite{rfc7567}.
\section{Conclusions}
\label{sec:conclusion}

We presented \name, a programmable packet scheduler that approximates the behavior of PIFO queues on existing data planes. Since \name can \emph{not} drop, nor modify the order of already-enqueued packets (as PIFO queues do), it estimates the expected distribution of incoming packet ranks, it predicts the admission and scheduling behaviors that the PIFO queue would follow, and it executes them \emph{at enqueue}. Our evaluation shows that \name is practical: it closely approximates PIFO behaviors and outperforms the state-of-the-art approaches.

\clearpage
\balance
\bibliographystyle{plain}
\bibliography{reference}

\clearpage
\nobalance
\appendix
\section{Appendix}
\label{sec:appendix}

\subsection{Theoretical analysis of \name}
\label{sec:theoretical analysis}

In the following, we show that, under certain conditions, the departure rate for \emph{all} packet ranks in \name is the same as for a PIFO queue. Moreover, under these conditions, there is only a small difference between the sets of packets forwarded by PIFO and \name.

\smallskip

Turning to the technical details, let the set of packets forwarded (up to time $t$) by PIFO and \name be $\text{PIFO}(t)$ and $\name(t)$, respectively. Then, to measure the difference in drops between \name and PIFO, we define:
$$ \Delta(t) = \frac{|\text{PIFO}(t) \setminus \name(t)| + |\name(t) \setminus \text{PIFO}(t)|}{|\text{PIFO}(t) + \name(t)|} .$$

We have $\Delta(t) \in [0,1]$, where a small value of $\Delta(t)$ indicates a small difference between \name and PIFO.
In the following, we denote the the maximal and minimal rank probabilities with $\delta_+ :=\max_i p(i)$ and $\delta_- := \min_i p(i)$.

\begin{theorem} Assume that the window size $|\mathcal{W}|$, buffer spaces $B_1, \dots, B_n$, and the number of arrived packets $T$ tend to infinity. Furthermore, assume that the maximal and minimal rank probabilities $\delta_+ $ and $\delta_-$ are bounded between two positive constants.
	We denote the ratio of the outgoing and incoming packet rate by $v$, and suppose $v<1$ (otherwise, both PIFO and \name behave like a FIFO).
	We claim that the difference between the drops of PIFO and \name is at most $\delta_+ $, i.e., $\Delta(T)_{T\to \infty}\le \delta_+$. Moreover, for each packet rank, the admittance rate of \name is identical to the admittance rate of PIFO.
\end{theorem}

\textbf{Proof:}
Since the window size, $|\mathcal{W}|$, is considered very large, the empirical rank distribution in $\mathcal{W}$ tends to the real packet rank distribution. 
In other words, after waiting a long time, we know rank probabilities with high precision, that is $\left|p(i) - p_{\mathcal{W}}(i)\right| \xrightarrow[T\ge |\mathcal{W}|]{|\mathcal{W}|\to \infty} 0$.
Thus, empirical quantiles, $\mathcal{W}.quantile (i)$, tend to the quantiles according to the real distribution, i.e., $\mathcal{W}.quantile (i) \to \sum_{i=1}^r p_i$.

Intuitively, since the buffer space $B$ is very large, the relative queue occupancy $b/B$ changes smoothly over time. More precisely, 
let $b(t)$ denote the queue occupancy after the arrival of the $t$\textsuperscript{th} packet (or, for short, `at time $t$'), and
let $q_n(t) =  \frac{1}{1-k} \frac{B-b(t)}{B} $ denote the highest queue bound at time $t$. 
At time $T$, we have queue bound $q_n(T)$ as the admission bound.
Let $r_T$ be the maximum rank such that $\mathcal{W}.quantile(r_T) \le q_n(T)$.
This means that the ratio of the admitted packets is $\sum_{i=0}^{r_T} p(i)$
Thus, after the arrival of the next packet, $\mathbb{E} (b(T+1) -b(T))= \sum_{i=0}^{r_T} p(i) - v$ (recall that, for every incoming packet, the number of drained packets is $v$ on average). 
This means the following.

\begin{enumerate}
	\item If $\sum_{i=0}^{r_T} p(i)>v$, the queue occupancy likely increases, ultimately triggering a drop in $q_n$ and in the rate of admitted packets.
	\item If $\sum_{i=0}^{r_T} p(i)<v$, the occupancy likely decreases, triggering a rise in $q_n$ and in the rate of admitted packets.
	\item Finally, in the event of $\sum_{i=0}^{r_T} p(i)=v$, the queue occupancy makes a motion very similar to the one-dimensional random walk, eventually, after a while likely triggering $q_n$ to either drop or rise for a short time period, before bouncing back to $r_T$.
\end{enumerate}
We note that, since the buffer spaces are considered to be very large, and the minimum rank probability $\delta_-$ is lower bounded by a positive constant, these events happen with probability $1$ based on the law of large numbers. Furthermore, in case $3$, $q_n(T+t)=r_T$ for any $t\ge0$ with probability $1$.

This also means that, in case $3$, $\Delta(T) \xrightarrow{T \to \infty} 0$, since after a while PIFO and \name forward the same packets with probability $1$. In cases $1$ and $2$, there is a single rank `on the border' that either gets forwarded or dropped by chance both in PIFO and \name; thus, in these cases, $\Delta(T)_{T \to \infty} \le \delta_+$. Note that the overall forwarding rate of this rank (and thus of all ranks) is the same for both PIFO and \name.

An alternative \textit{intuitive} reasoning supporting the statement that the forwarding rates coincide for PIFO and \name is the following. In both cases, there are three classes of ranks: i) small ranks that always are forwarded, ii) large ranks that always are rejected, and finally iii) a borderline rank $r^*$ that is either forwarded or rejected by chance. Since draining is continuous both for PIFO and \name, the leftover bandwidth after the small-ranked packets is given to the borderline rank, $r^*$, as it is the only choice, again both for PIFO and \name. \hfill $\square$

\end{document}